\documentclass{JFM-FLM_Au}

\usepackage{bm}
\usepackage{physics}
\usepackage{tikz-cd}
\usepackage{subcaption}
\usepackage[export]{adjustbox}
\usepackage{multirow}
\usepackage{comment}
\usepackage{pifont}
\usepackage[dvipsnames]{xcolor}

\DeclareMathOperator{\vflat}{vec}

\DeclareMathOperator{\sym}{Sym}
\newcommand{\field}[1]{\mathbb{#1}}
\newcommand{\R}{\field{R}}

\newcommand{\oth}{\mathrm{O}(3)}
\newcommand{\cg}{\otimes_{\mathrm{cg}}}
\newcommand{\kr}{\otimes_{\mathrm{kr}}}

\newcommand{\mata}[1]{\frac{\bar{\mathrm{D}}#1}{\bar{\mathrm{D}}t}}
\newcommand{\mc}[1]{\mathcal{#1}}
\newcommand{\avg}[1]{\langle#1\rangle}
\newcommand{\Avg}[1]{\left\langle#1\right\rangle}
\newcommand{\tens}[1]{\mathsfbi{#1}}
\newcommand{\tgreek}[1]{\bm{#1}}
\newcommand{\rst}{\tens{R}}
\newcommand{\rs}{R}
\newcommand{\cmark}{\textcolor{ForestGreen}{\ding{51}}}
\newcommand{\xmark}{\textcolor{BrickRed}{\ding{55}}}
\renewcommand{\b}[1]{\bm{#1}}
\renewcommand{\c}{\texttt}

\newtheorem{define}{Definition}[section]

\makeatletter
\renewenvironment{table}[1][tp]
  {\@float{table}[#1]\fstyle@table}
  {\end@float}
\renewenvironment{table*}[1][tp]
  {\@dblfloat{table}[#1]\fstyle@table}
  {\end@dblfloat}
\makeatother

\lefttitle{A. Miller, S. Kommalapati, R. Moser, and P. Koumoutsakos}
\righttitle{Journal of Fluid Mechanics}

\title{Structure tensor Reynolds-averaged Navier-Stokes turbulence models with equivariant neural networks}

\author{Aaron Miller\aff{1}, Sahil Kommalapati\aff{2, 3}, Robert Moser\aff{2, 3}, \and Petros Koumoutsakos\aff{1}}

\affiliation{
    \aff{1}Computational Science and Engineering Laboratory, Harvard University, MA 02138, USA
    \aff{2}Oden Institute for Computational Engineering and Sciences, The University of Texas at Austin, TX 78712, USA
    \aff{3}Walker Department of Mechanical Engineering, The University of Texas at Austin, TX 78712, USA
}

\corresau{Aaron Miller, \email{aaronmiller@g.harvard.edu}}

\begin{document}
\maketitle

\begin{abstract}
    Accurate and generalizable Reynolds-averaged Navier-Stokes (RANS) models for turbulent flows rely on effective closures, but currently available closures are notoriously unreliable. Kassinos \textit{et al.} (J. Fluid Mechanics, \textbf{428}, pp. 213-248, 2001) hypothesized that this unreliability of RANS models was due to an insufficient description of the statistical state of the turbulence and proposed a set of structure tensors as a candidate for a sufficiently rich description. To test this hypothesis for the rapid pressure-strain term, we introduce tensor-based, symmetry aware closures in terms of the structure tensors using equivariant neural networks (ENNs), and present an algorithm for enforcing algebraic contraction relations among tensor components. Using data from rapid distortion theory, experiments show that such ENNs can effectively learn relationships involving high-order tensors. The resulting ENN structure tensor models are orders of magnitude more accurate than existing models for the rapid pressure-strain correlation, effectively validating the Kassinos \textit{et al.} hypothesis for this term. Results show that ENNs provide a physically consistent alternative to classical tensor basis models, enabling end-to-end learning of unclosed terms in RANS and other tensor modeling domains, and rapid exploration of model dependencies.
\end{abstract}

\begin{keywords}
    % Authors should not enter keywords on the manuscript, as these must be chosen by the author during the online submission process and will then be added during the typesetting process (see \href{https://www.cambridge.org/core/journals/journal-of-fluid-mechanics/information/list-of-keywords}{Keyword PDF} for the full list).  Other classifications will be added at the same time.
\end{keywords}

% {\bf MSC Codes }  {\it(Optional)} Please enter your MSC Codes here

\section{Introduction}

Reynolds-averaged Navier-Stokes (RANS) is the most commonly used formulation for computational simulation of turbulent flows in applications such as automobile and aircraft aerodynamics. Its lower cost, compared to higher-fidelity methods, such as Direct Numerical Simulations (DNS) and Large Eddy Simulation (LES), makes RANS a practical and efficient choice, providing statistically averaged flow information rather than resolving turbulent fluctuations. In RANS, flow quantities are decomposed into mean and fluctuating parts, and the governing Navier-Stokes equations are used to develop equations for the mean velocity. These equations involve the Reynolds stress tensor that needs to be modeled, leading to the RANS closure problem. Throughout this work, the term \emph{closure} indicates a model of a particular flow variable appearing in the governing equations, in contrast to use of the term to indicate a correction added to remedy modeling or discretization error.

A popular approach for closing the RANS equations is to adopt the linear turbulent (or eddy) viscosity hypothesis, which models the Reynolds stress tensor as proportional to the mean strain rate tensor, analogous to the viscous stress relation for a Newtonian fluid. Although computationally efficient and simple to implement, the linear turbulent viscosity assumption does not hold for many important flow conditions, particularly those that include curved geometries or separation \citep{pope_turbulent_2000}. Nonlinear turbulent viscosity models, which feature higher-order products of the mean strain rate and rotation rate tensors, have also been proposed \citep{pope_more_1975}. An alternative closure approach that avoids the turbulent viscosity assumption involves solving transport equations for the Reynolds stress tensor directly. As Reynolds stresses are the second moments of velocity, these are often referred to as second-order closure models \citep{lumley_computational_1979}, or second moment closures.

Many models to close the Reynolds stress transport equations have been proposed \citep{speziale1987second,launder_second-moment_1989,speziale1991modelling,so1991second,durbin1993reynolds}, many of which were built on the models of Launder, Reece \& Rodi \citep{launder_progress_1975}. But, virtually all practical applications of RANS use eddy viscosity models, despite the fact that Reynolds stress transport models should be higher fidelity and more broadly applicable \citep{durbin2011statistical}. Among the reasons is that, with currently available closure models, the Reynolds stress transport equations are more expensive and more difficult to solve, while often providing no significant improvement in accuracy over eddy viscosity models. One of the most challenging terms to model is the rapid pressure-strain correlation (rapid term). \citet{kassinos_one-point_2001} hypothesized that the reason the rapid term was difficult to model is that the Reynolds stress tensor did not adequately describe the anisotropy of the turbulence, so that a model depending on the Reynolds stress could not represent all the effects of turbulent anisotropy. They proposed a richer set of tensors to describe turbulence anisotropy that they called structure tensors, and hypothesized that they were sufficient to accurately model the rapid pressure-strain term.  It is also plausible that the structure tensors are a sufficiently rich representation that they enable accurate models for unclosed terms in their own transport equations.

Kassinos and Reynolds were able to show that a linear structure tensor model for the rapid term in the Reynolds stress transport equation could represent the rapid pressure-strain term very accurately in the case of pure rotation. However, with the computational tools available at the time, they were unable to consider general nonlinear models or more general mean gradients. In this paper, we revisit the Kassinos and Reynolds hypothesis to develop nonlinear structure-tensor-based models for the rapid terms for use with general homogeneous mean velocity gradients. This is pursued in the context of rapid distortion theory (RDT) \citep{hunt1990rapid},  since in RDT the rapid terms are the only ones requiring modeling to close the Reynolds stress or structure tensor transport equations. Further, solving the RDT equations numerically is computationally inexpensive, so that large quantities of data can be easily generated.

In general, closing the Reynolds stress or structure function transport equations requires representing unclosed terms as tensor-valued functions of tensor arguments. The structure of these functions is constrained by the transformation properties of the tensors under the $\oth$ symmetry group, which ensures that the modeled physics is invariant to the coordinate system in which the model is expressed. This notion is fundamental in turbulence modeling \citep{pope_turbulent_2000,lumley_stochastic_2007}, and is further emphasized with the rise of data-driven approaches \citep{spalart_old-fashioned_2023,duraisamy_turbulence_2019}. 

In this paper, we construct symmetry consistent RANS closure models by representing tensor functions as equivariant neural networks (ENNs). While previous approaches require a priori tensor analysis to ensure physical validity, ENNs can enforce several symmetries as hard constraints through architecture design. Here, an algorithm is presented that broadens the types of constraints that ENNs can enforce to enable their application to RANS modeling. The primary contributions of the paper are:

\begin{itemize}
    \item Introduction of a novel algorithm and corresponding ENN layer that exactly enforces linear tensor relationships, extending the symmetry preserving properties of ENNs to a new class of constraints beyond equivariance and index permutation symmetries. We note that,  although applied here to RANS, the method is general and can be used in other domains without modification.
    \item Development of the first general nonlinear model for pressure-strain related terms for RANS models depending on the structure tensors of Kassinos and Reynolds \citep{kassinos_one-point_2001}, in addition to a broader group-theoretic analysis of the turbulent structure tensors.
    \item Evaluation of ENN structure tensor models based on new rapid distortion theory data, which demonstrates that the structure tensor models for the rapid terms have orders of magnitude lower errors than currently available models.
\end{itemize}

In Sec.~\ref{sec:related}, we review previous approaches for the representation of equivariant tensor functions in RANS modeling, as well as the development of ENNs in the broader machine learning literature. Sec.~\ref{sec:bg} describes the problem setting and Sec.~\ref{sec:methods} the representation theory of tensors, how it informs the design of the ENN structure and the proposed algorithm to enforce the linear tensor constraints. Sec.~\ref{sec:results} describes data generation and results. Finally, Sec.~\ref{sec:discussion} concludes with a discussion and outlook.

\section{Related Work}\label{sec:related}

\subsection{Invariant theory for tensor function representation}\label{sec:tbm-bg}

The dominant approach in fluid mechanics to build equivariant tensor functions was initiated by \citet{robertson_invariant_1940} and applies the theory of invariants to represent tensor functions as a linear combination of basis tensors \citep{lumley_computational_1979,sarkar_simple_1989,chung_nonlinear_1995,ling_reynolds_2016,kaandorp_data-driven_2020,cai_revisiting_2024,spencer_theory_1958}. The expansion coefficients are, in general, functions of tensor invariants. Although the form of these coefficients must be determined by other means, this approach satisfies the appropriate symmetry properties while reducing the modeling problem to the determination of scalar functions of scalar arguments. This approach is here referred to as the \textit{tensor basis method} (TBM).

Analytical methods for determining the coefficient functions include treating them as constants, forming expressions from Taylor expansions \citep{chung_nonlinear_1995}, or deriving other non-polynomial forms \citep{lumley_computational_1979}. The drawback to such approaches is that they rely on a priori assumptions or require truncation. Alternatively, the tensor basis neural network (TBNN), introduced by \citet{ling_reynolds_2016} and applied in \citep{zhang_ensemble_2022,lennon_scientific_2023,parmar_generalized_2020}, among others, uses neural networks to parameterize the coefficient functions. The universal approximation property of neural networks implies that this construction remains theoretically general, and since the coefficient functions are scalar functions of scalars, any network architecture can be used without violating symmetry constraints.

The TBM has been used successfully for several turbulence modeling problems, for example predicting the Reynolds stress anisotropy and improving mean velocity predictions in square duct, wavy wall, curved backward-facing step, and right-angled backward facing step geometries \citep{ling_reynolds_2016,kaandorp_data-driven_2020}, as well as for modeling tasks in other areas of fluid mechanics, for example learning rheological constitutive equations \citep{lennon_scientific_2023,sunol_learning_2025}. However, it is limited by the need to derive the tensor basis, which determines both the number of coefficient functions and their arguments. Deriving a minimal basis and the associated invariants is a challenging task. The size of the minimal integrity basis grows exponentially with the number of tensor arguments, and the results are only available for a limited number of special cases \citep{smith_isotropic_1971,pennisi_third_1992,spencer_theory_1958,prakash_invariant_2022}. Moreover, a newly derived basis is necessary when the model dependencies change. This motivates the search for alternative modeling procedures that satisfy the relevant symmetry constraints and maintain generality but do not require a priori derivation of basis tensors and their invariants.

\subsection{Equivariant neural networks}

The architectures of ENNs are distinguished by the way in which they handle geometric attributes \citep{duval_hitchhikers_2024}. Some approaches use only scalar features, such as norms or triplet angles, as input to the network instead of higher-order tensors \citep{schutt_schnet_2017,gasteiger_directional_2022}. While this guarantees invariance to Euclidean transformations, it comes at the cost of expressivity. Moreover,  pre-computation of the derived scalars becomes increasingly complex with increasing numbers and tensor orders of the inputs. Alternatively, equivariant architectures aim to retain full geometric information by operating on tensors directly. The developments that most inform the present work include those in \citep{thomas_tensor_2018,weiler_3d_2018,brandstetter_geometric_2021,kondor_clebsch-gordan_2018}, which make use of the notion of irreducible representations. Working directly with the fundamental building blocks of group representations provides a framework for generalizing equivariant operations to higher-order tensors, a task that is particularly relevant to RANS applications.
ENNs have been used in particular to obtain state-of-the-art property predictions in molecular dynamics simulations \citep{jumper_highly_2021,batzner_e3-equivariant_2022,zitnick_introduction_2020,dauparas_robust_2022}. These results have demonstrated the efficacy of operating directly on geometric attributes. The TBM, discussed above, applies machine learning only to the \textit{invariant} parts of the problem, relying on analytical treatment of the \textit{equivariant} aspects offline.

An equivariant closure modeling strategy that avoids offline basis derivation is the vector cloud neural network (VCNN-e) \citep{zhou_frame-independent_2022,han_equivariant_2023}, which maps a cloud of local flow or geometry features to the Reynolds stresses via an invariant embedding and an equivariant output map. However, it only considers vectorial features and does not provide a framework for enforcing general index permutation symmetries or prescribed linear constraints. Another approach targeting equivariant tensorial constitutive relations is RotEqNet \citep{gao_roteqnet_2022}, which uses position standardization to map rotated inputs to a canonical orientation. For higher-order tensors, contractions are used to compute a rotation into the standard position. A model is trained on the standardized inputs, and the predictions are rotated back to the original frame for inference. This method is restricted to fully symmetric tensors and guarantees exact equivariance only when the model in the standardized frame is learned perfectly.

The work of \citet{kaszuba_implicit_2025} explores the use of irreducible representation-based equivariant networks in RANS but differs from the present study in several key respects. First, the model in \citet{kaszuba_implicit_2025} is trained to learn single-injection corrections based on the converged output of a particular RANS model. This limits the applicability of the model to inputs associated with the RANS solver used to generate training data. In contrast, this work learns constitutive relations between terms in the governing equations, and hence completes a RANS model rather than corrects it. Furthermore, the present work targets Reynolds stress models and does not restrict the focus to steady problems.

Equivariant networks have also been applied to problems in fluid dynamics beyond turbulence modeling. One line of research extends the mesh-based graph neural network (GNN) surrogate framework introduced in \citep{pfaff_learning_2021}. For example, \citet{toshev_learning_2023} applies the Steerable Equivariant Graph Neural Network architecture to Lagrangian fluid mechanics, providing a direct comparison to non-equivariant surrogates. Related GNN approaches employ an encode-process-decode strategy, but modify the treatment of vector features to enforce equivariance, either by transforming them into local strain eigenbases \citep{list_rotational_2025} or by projecting them onto graph edges \citep{lino_multi-scale_2022}.

\subsection{Integrating physics constraints in  machine learning models}\label{sec:prior-constraints}

Integrating physical priors into machine learning models aims to improve their data efficiency and predictive accuracy. To our knowledge, the first work in this field integrated wall shear stresses and pressure gradients to improve the neural network forecast of near-wall flow in turbulent channel flows \citep{milano_neural_2002}. Physics-informed machine learning models such as PINNS \citep{raissi_physics-informed_2019,karniadakis_physics-informed_2021,wang_incorporating_2021} embed physics constraints by regularizing the loss function with a term that penalizes deviations from known PDE governing equations. In order to achieve equivariance, a straightforward approach is data augmentation, in which symmetry operations are applied to the training data \citep{chen_group-theoretic_nodate}. However, at least for continuous symmetry groups, it is impossible to generate data for every possible symmetry transformation, and there is no guarantee that the model will satisfy equivariance.

Alternatively, \textit{hard constraints} reduce the search space for optimization and ensure physical consistency by construction. Examples of this approach include \citep{liu_harnessing_2024,richter-powell_neural_2022,chalapathi_scaling_2024}. The tools described here also fall into the hard constraint category. ENNs are equivariant by construction and can be designed to respect index permutation symmetries, such as for symmetric tensors. In this paper, a method is developed for enforcing another class of tensor relationships, namely ``einsum"-style linear relations, as hard constraints in ENNs. Einsum is a generalized Einstein summation notation to represent arbitrary linear maps and relations that can extend beyond standard tensor contractions and has found extensive applications in machine learning. An einsum-like domain-specific language is the basis for the Tensor Comprehensions \citep{vasilache_tensor_2018} library in PyTorch, which automatically generates GPU code and auto-tunes it for specific input sizes.

\section{Reynolds-Averaged Navier-Stokes (RANS) Modeling}\label{sec:bg}

In RANS models, the Reynolds decomposition is applied to the velocity $\b{u}=\b{U}+\b{u}'$ and pressure $p=P+p'$ where $\b{U}=\avg{\b{u}}$ and $P=\avg{p}$ are the expected values of the velocity and pressure, respectively. Substituting this decomposition into the incompressible Navier-Stokes equations and averaging yields the incompressible Reynolds-averaged Navier-Stokes equations for the mean velocity, written here using Cartesian tensor notation and the Einstein summation convention:
\begin{equation}
  \mata{U_i} =-\frac1{\rho}\pdv{P}{x_i}+\nu\frac{\partial^2U_i}{\partial x_j \partial x_j}-\pdv{\avg{u'_iu'_j}}{x_j},\qquad \pdv{U_i}{x_i}=0,
  \label{eq:MeanEq}
\end{equation}
where $\mata{}\equiv\pdv{}{t}+U_k\pdv{}{x_k}$ is the mean convective time derivative.
As is well known, the primary challenge in RANS is specifying a model for the Reynolds stress tensor $R_{ij}\equiv\avg{u'_iu'_j}$.

In Reynolds stress transport modeling, which is the approach of interest here, the Navier-Stokes equations and the Reynolds averaged equations (Eq.~\ref{eq:MeanEq}) are manipulated to determine an exact equation for the Reynolds stress:
\begin{equation}
  \mata{R_{ij}} = \mc{P}_{ij} +\Pi_{ij} -\epsilon_{ij} + \pdv{T_{ijk}}{x_k},\label{eq:RSeq}
\end{equation}
where
\begin{align*}
  \mc{P}_{ij} &\equiv -R_{ik}\pdv{U_j}{x_k}-R_{jk}\pdv{U_i}{x_k},\\
  \Pi_{ij} &\equiv \frac1{\rho}\left\langle p'\left(\pdv{u'_i}{x_j}+\pdv{u'_j}{x_i}\right)\right\rangle,\\
  T_{ijk}  &\equiv -\avg{u'_iu'_ju'_k} - \frac1{\rho}\left(\avg{u'_ip'}\delta_{jk}+\avg{u'_jp'}\delta_{ik}\right) +\nu\pdv{R_{ij}}{x_k},\\
  \epsilon_{ij} &\equiv 2\nu\left\langle\pdv{u'_i}{x_k}\pdv{u'_j}{x_k}\right\rangle,
\end{align*}
are the production, pressure-strain, transport flux, and dissipation terms, respectively. The transport flux $T_{ijk}$ includes transport due to turbulent fluctuations, pressure-velocity correlations, and viscosity. The last three terms on the right-hand side of Eq.~\ref{eq:RSeq} require closure models. Of particular interest here will be modeling the pressure-strain term $\Pi_{ij}$, which requires analyzing the fluctuating pressure.

\subsection{The rapid pressure-strain term}

The pressure fluctuation $p'$ appearing in the pressure-strain term and the pressure-velocity correlation is determined as the solution of a Poisson equation
\begin{equation}
    \frac{1}{\rho}\laplacian p' = -2G_{ij}\pdv{u'_j}{x_i} - \frac{\partial (u_i'u_j'-\rs_{ij})}{\partial x_i\partial x_j},\label{eq:poisson}
\end{equation}
along with appropriate boundary conditions on $p'$ (generally periodic or Neumann). Here, $G_{ij}=\pdv{U_i}{x_j}$ is the mean velocity gradient tensor. Note that the first term on the right-hand side of Eq.~\ref{eq:poisson} depends linearly on the mean velocity gradient, while the second term depends only on the fluctuating velocity. In modeling the pressure-strain term, it is useful to decompose $p'$ into the part due to the first term and the part due to the second term, since they have different dependencies. To this end, pressure is decomposed as $p'=p^{(\mathrm{r})}+p^{(\mathrm{s})}+p^{(\mathrm{h})}$ with:
\begin{align*}
    \frac{1}{\rho}\laplacian p^{(\mathrm{r})} &= -2G_{ij}\pdv{u_j'}{x_i},\\
    \frac{1}{\rho}\laplacian p^{(\mathrm{s})} &= -\frac{\partial (u_i'u_j'-\rs_{ij})}{\partial x_i\partial x_j},\\
    \frac{1}{\rho}\laplacian p^{(\mathrm{h})} &= 0,
\end{align*}
where $p^{(\mathrm{r})}$ is the ``rapid pressure'' since it responds immediately to a change in the mean velocity gradient, $p^{(\mathrm{s})}$ is the slow pressure, and $p^{(\mathrm{h})}$  is the homogeneous or Stokes pressure. Periodic or homogeneous Neumann boundary conditions are generally applied to $p^{(\mathrm{r})}$ and $p^{(\mathrm{s})}$, whereas $p^{(\mathrm{h})}$ has the same boundary conditions as $p'$. 
The corresponding pressure-strain tensors are then
\begin{equation}\label{eq:pressure-strain-def}
    \Pi_{ij}^{(\cdot)} \equiv \Avg{\frac{p^{(\mathrm{\cdot})}}{\rho}\left(\pdv{u'_i}{x_j} + \pdv{u'_j}{x_i}\right)},
\end{equation}
where the dot in $(\cdot)$ is one of $\{\mathrm{r}, \mathrm{s}, \mathrm{h}\}$.

A particularly challenging term to model has been the rapid pressure-strain term $\Pi_{ij}^{\mathrm{(r)}}$. A number of different models have been proposed for the rapid term over the years \citep{launder_progress_1975,speziale1987second,speziale1991modelling,johansson1994modelling,girimaji2000pressure,mishra2010pressure,panda2020review}, with limited success. In the case of homogeneous turbulence subjected to a mean velocity gradient, which we will consider here, the transport terms drop out of Eq.~\ref{eq:RSeq}, the Stokes pressure is zero, and the rapid term can be written
\begin{equation}
\Pi^{\mathrm{(r)}}_{ij}=2G_{lk}(M_{ikjl}+M_{jkil}).\label{eq:rapidPi}
\end{equation}
Here, the fourth order tensor $\tens{M}$ can be defined in terms of the velocity spectrum tensor $\Phi_{ij}(\b{\kappa})$ as
\begin{equation}
M_{ijkl}=\int\frac{\kappa_k\kappa_l}{\kappa^2}\Phi_{ij}\,\dd{\b{\kappa}}.\label{eq:M-def}
\end{equation}
Thus, in this case, modeling the rapid term amounts to modeling $\tens{M}$.

\subsection[M models and structure tensors]{$\tens{M}$ models and structure tensors} 

Kassinos \& Reynolds \citep{kassinos_one-point_2001} hypothesized that the reason for the limited success of rapid pressure-strain models that are posed in terms of the Reynolds stress is that the Reynolds stress is not a sufficiently rich description of the statistical state of the turbulence. Particularly, the Reynolds stress characterizes the anisotropy of the fluctuating velocity components, but it is insensitive to anisotropies in length scale. Kassinos \& Reynolds proposed a set of structure tensors that provide a more complete characterization of turbulence anisotropy. They included the Reynolds stress ($R_{ij}$), the dimensionality ($D_{ij}$) that characterizes length scale anisotropy, circulicity ($F_{ij}$) that characterizes anisotropy in rotational motion, and stropholysis ($Q^*_{ijk}$) that characterizes symmetry breaking. The properties of these tensors are described in detail in \citep{kassinos_structure-based_1995,kassinos_one-point_2001}; here we recount those properties important for the ENN modeling explored here.

First, all four of the above tensors are fully symmetric (i.e. they are invariant to arbitrary index permutations). Further all traces of $Q^*_{ijk}$ are zero, and $R_{ii}=D_{ii}=F_{ii}=q^2$, where $q^2\equiv R_{ii}$ is the variance of the fluctuating velocity magnitude. For homogeneous turbulence, the second-order tensors are related by
\begin{equation*}
R_{ij}+D_{ij}+F_{ij}=\delta_{ij}q^2,
\end{equation*}
so one only needs to consider the Reynolds stress and dimensionality
tensors. The stropholysis is the symmetrization of a more general
third-order tensor $Q_{ijk}=\epsilon_{ipq}M_{jqpk}$, which, for
homogeneous turbulence, is also trace-free. For homogeneous turbulence, $Q_{ijk}$ decomposes as follows:
\begin{equation}
Q_{ijk}=\frac16(q^2\epsilon_{ijk}+2\epsilon_{ikm}R_{mj}+2\epsilon_{jim}D_{mk}+2\epsilon_{kjm}F_{mi})+Q^*_{ijk},\label{eq:Qdecomp}
\end{equation}
and the second-order tensors can be expressed in terms of $Q_{ijk}$
\begin{equation}
R_{ij}=\epsilon_{imp}Q_{mjp},\qquad D_{ij}=\epsilon_{imp}Q_{pmj},\qquad F_{ij}=\epsilon_{imp}Q_{jpm}.\label{eq:Qgen}
\end{equation}
These relations imply that for homogeneous turbulence, it is equivalent to use $R_{ij}$, $D_{ij}$, and $Q^*_{ijk}$ as descriptors of turbulence anisotropy, or to use $Q_{ijk}$.

To use the structure tensors as a basis for turbulence modeling, it will be necessary to solve an evolution equation for them. For the current purpose of modeling the rapid pressure-strain term, it will be sufficient to consider modeling in the simpler context of rapid distortion theory. In RDT, one assumes that the magnitude of the mean velocity gradient is much larger than the inverse eddy turn-over time of the turbulence (estimated at $q^2/\epsilon$, where $\epsilon$ is the rate of kinetic energy dissipation). In this limit, terms in the Navier-Stokes equations that are nonlinear in the velocity fluctuations are negligible, as are the viscous terms. The result is that the only terms remaining on the right-hand side of the Reynolds stress transport equation (Eq.~\ref{eq:RSeq}) are the production term and the rapid term. There is a similar simplification of the other structure tensor evolution equations. Due to Eqs.~\ref{eq:Qdecomp} and \ref{eq:Qgen}, it is sufficient to consider the evolution equation for $Q$, since the other structure tensors can be found from it. In the RDT limit, the evolution equation for $Q_{ijk}$ is

\begin{equation}\label{eq:rdt-q}
\begin{split}
\dv{Q_{ijk}}{t} = &-G_{mk}Q_{ijm}-G_{jm}Q_{imk}-\epsilon_{its}(G_{sm}M_{jmtk}+G_{mt}M_{jsmk})\\
&-\Omega_t(\delta_{it}\epsilon_{jpq}Q_{qpk}-L_{itjk}-M_{itjk})
+2S_{tm}(J_{ijktm}+J_{itmjk}),
\end{split}
\end{equation}
where $\Omega_i=\epsilon_{ijk}G_{kj}$ is the mean vorticity, $S_{ij}=\frac12(G_{ij}+G_{ji})$ is the mean strain rate, and
\begin{equation}\label{eq:LJ-def}
L_{ijkl}=\int\frac{\kappa_i\kappa_j\kappa_k\kappa_l}{\kappa^4}\Phi_{mm}(\b{\kappa})\,\dd{\b{\kappa}},
\qquad
J_{ijrpq}=\epsilon_{its}\int\frac{\kappa_t\kappa_r\kappa_p\kappa_q}{\kappa^4}\Phi_{sj}(\b{\kappa})\,\dd{\b{\kappa}}.
\end{equation}
The terms in the first line of Eq.~\ref{eq:rdt-q} are the analog of
production terms, and the terms on the second line are analogous to the rapid pressure-strain term. Note that in addition to the tensor $M_{ijkl}$ that appears in the rapid term in the Reynolds stress equation, there are two new tensors $L_{iklm}$ and $J_{ijkmn}$ that must also be modeled in terms of $Q_{ijk}$ (or equivalently in terms of $R_{ij}$, $D_{ij}$ and $Q^*_{ijk}$). Modeling these tensors using ENNs is the objective in the remainder of this paper.

\section{Methodology}\label{sec:methods}

The theoretical foundations of ENNs draw extensively on group and representation theory, as these closely related fields provide a natural way to formalize symmetry relationships. In this paper, relevant definitions and results are recalled as needed, and further details can be found in \citep{dresselhaus_group_2008,georgi_lie_2018}.

\subsection{Cartesian tensors and spherical tensors}\label{sec:tensors}

Flow variables in RANS models have geometric attributes in a three-dimensional Euclidean space. Abstractly, physical vector quantities such as velocity and force are simply vectors (tensors of order one) in this space. Higher-order tensor quantities (e.g. stress) define linear mappings from lower-order tensor quantities to other lower-order tensors. Supposing the use of a Cartesian basis for the Euclidean space, tensors are represented as ``Cartesian tensors,'' which are arrays of components that transform under rotation and reflection to ensure equivariance of the mappings (see Sec.~\ref{sec:EquiVar}) to rotations of the Cartesian basis.  

\begin{define}[Cartesian tensor]\label{def:ct}
    A generalized Cartesian tensor of order $n$ and signature $s\in\{0, 1\}$ is a quantity defined by its $3^n$ components in a Cartesian coordinate system, which transform according to the following rule under an orthogonal coordinate transformation described by the matrix $\tens{R} \in \R^{3\times 3}$:
    \begin{equation}\label{eq:ct-def}
        T_{i_1i_2\cdots i_n}' = (\det(\tens{R}))^sR_{i_1j_1}R_{i_2j_2}\cdots R_{i_nj_n}T_{j_1j_2\cdots j_n}.
    \end{equation}
\end{define}

The signature $s$ determines a tensor's parity, or its behavior under reflection. It extends the strict definition of a Cartesian tensor to describe quantities that behave differently under improper rotations (a combination of a rotation and a reflection), but are nonetheless physically relevant. For example, under an improper rotation $\tens{R}$, quantities such as velocity, force, and electric field transform as $T_i' = R_{ij}T_j$, while quantities such as angular momentum, torque, and magnetic field transform as $T_i' = -R_{ij}T_j$. Vectors and tensors with $s = 1$ are called pseudovectors and pseudotensors, respectively.

Many useful tools become available by viewing Def.~\ref{def:ct} from a group theoretic perspective. The orthogonal transformations $\{\tens{R}\in\R^{3\times 3} | \tens{R}^{\mathrm{T}}\tens{R} = \tens{I}\}$ form a group, namely the orthogonal group $\oth$, under matrix multiplication, and the collection of tensors of a given order and dimensionality constitutes a vector space. Hence, Eq.~\ref{eq:ct-def} essentially defines a Cartesian tensor as a collection of components that transform according to a particular representation of $\oth$, namely the tensor product representation.

Although this representation is an intuitive generalization of the familiar matrix-vector product for rotating a vector, it is reducible, meaning that it contains at least one nontrivial invariant subspace. It is desirable to work with irreducible representations, as this makes it clear which operations are allowable when interacting tensors in an equivariant neural network. For compact groups, any finite-dimensional representation is completely reducible, meaning it can be decomposed as a direct sum of irreducible representations via a change of basis. The bases that achieve this for $\oth$ are spherical bases, in which abstract tensors are represented as ``spherical tensors.''%\footnote{The more common setting in which spherical tensors appear is quantum mechanics, where there is the notion of a spherical tensor \textit{operator}. The definition presented here is formulated to mirror Def.~\ref{def:ct}, and although it is not the same as the quantum mechanical one, it represents the same core idea of components mixing within a single irreducible subspace of the rotation group.}.

\begin{define}[Spherical tensor]
    A spherical tensor of order $\ell$ and parity $p\in\{-1, 1\}$ is a set of $2\ell + 1$ components, denoted $T_m^{(\ell, p)}$, where $m$ takes on integer values $-\ell, -\ell + 1, \dots, \ell$, that transform under a coordinate transformation $\tens{R}\in\oth$ as%\footnote{The symmetric indexing used here, while different from the standard zero-based indexing $0, 1, \dots$, is standard in the equivariant machine learning literature and has its roots in quantum mechanics.}
    \begin{equation}\label{eq:sph-def}
        T_m'^{(\ell, p)} = p^{(1-\det(\tens{R}))/2}\sum_{m' = -\ell}^\ell \mathcal{D}^{(\ell)}_{mm'}(\hat{\tens{R}})T_{m'}^{(\ell, p)},
    \end{equation}
    where $\hat{\tens{R}} \equiv \det(\tens{R})\tens{R}$ is the pure rotational part of $\tens{R}$ and $\mathcal{D}^{(\ell)}_{mm'}(\hat{\tens{R}})$ are the matrix elements of the order-$\ell$ irreducible representation of the special orthogonal group $\mathrm{SO}(3)$. The $(2\ell + 1)\times (2\ell + 1)$ matrices $\tens{D}^{(\ell)}(\hat{\tens{R}})$ are known as the Wigner-D matrices. It is convenient to define%\footnote{A more explicit notation would include the parity, for example $\hat{\tens{D}}^{(\ell, p)}(\tens{R})$, but this is avoided for clarity. The hat notation is used instead to simultaneously reflect the connection to the Wigner-D matrices and mark the distinction that comes from considering $\oth$ instead of $\mathrm{SO}(3)$.}
    \begin{equation*}
        \hat{\tens{D}}^{(\ell)}(\tens{R}) \equiv p^{(1-\det(\tens{R}))/2}\tens{D}^{(\ell)}(\hat{\tens{R}}),
    \end{equation*}
    such that the above definition may also be written as
    \begin{equation*}
        T_m'^{(\ell, p)} = \sum_{m' = -\ell}^\ell \hat{\mathcal{D}}^{(\ell)}_{mm'}(\tens{R})T_{m'}^{(\ell, p)}. 
    \end{equation*}
\end{define}

\begin{figure}
    \centering
    \begin{equation*}
        % store the pmatrix in a box to measure its height
        \sbox0{
          $\begin{pmatrix}
            T_{00} \\
            T_{01} \\
            T_{02} \\
            T_{10} \\
            T_{11} \\
            T_{12} \\
            T_{20} \\
            T_{21} \\
            T_{22}
          \end{pmatrix}$
        }
        \underset{\raisebox{-2ex}{\text{\large$\tens{R}\kr\tens{R}$}}}{\adjincludegraphics[height=\the\ht0+\the\dp0, valign=c]{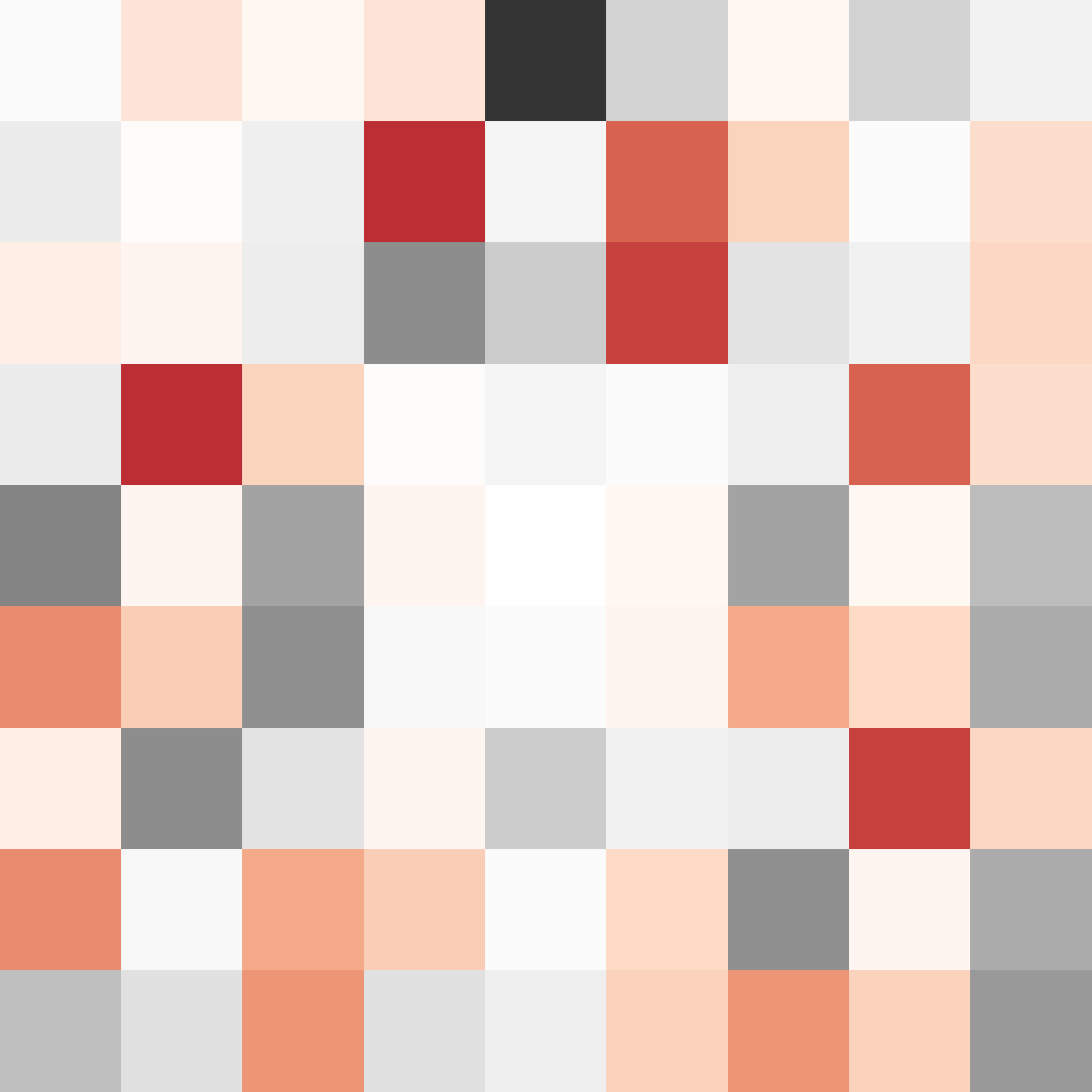}}
        \underset{\raisebox{-2ex}{\text{\large$\vflat\qty(\tens{T})$}}}{\begin{pmatrix}
            T_{00} \\
            T_{01} \\
            T_{02} \\
            T_{10} \\
            T_{11} \\
            T_{12} \\
            T_{20} \\
            T_{21} \\
            T_{22}
        \end{pmatrix}}
        \quad 
        \overset{\raisebox{1ex}{\text{\large$\tens{S}_\rho$}}}{\longrightarrow}
        \quad
        \underset{\raisebox{-2ex}{\text{\large$\bigoplus_{\ell=0}^2 \hat{\tens{D}}^{(\ell)}(\tens{R})$}}}{\adjincludegraphics[height=\the\ht0+\the\dp0, valign=c]{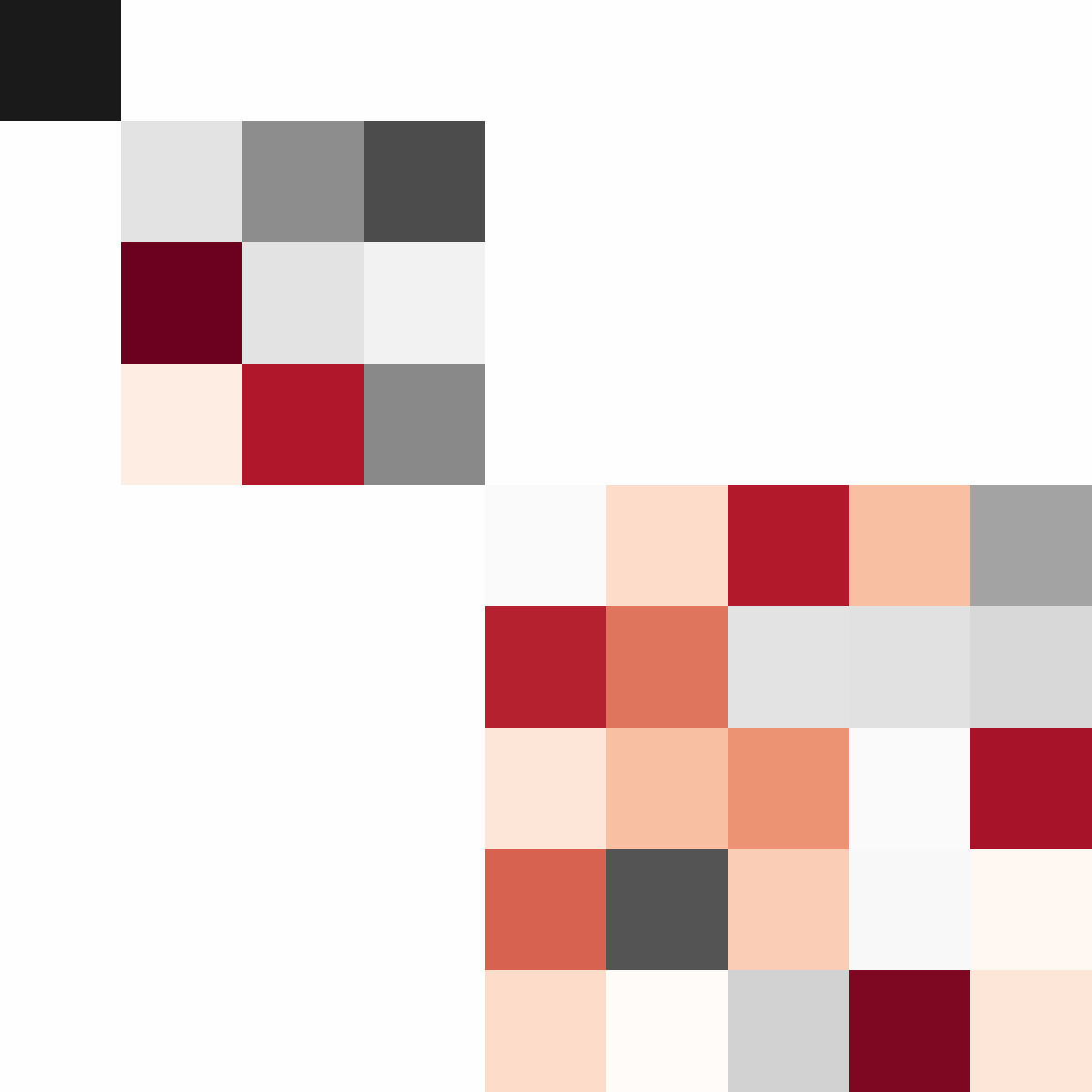}}
        \ 
        \underset{\raisebox{-2ex}{\text{\large$\bigoplus_{\ell=0}^2\tens{T}^{(\ell, p)}$}}}{\begin{pmatrix}
            T_{00} + T_{11} + T_{22} \\
            T_{12} - T_{21} \\
            T_{20} - T_{02} \\
            T_{01} - T_{10} \\
            2T_{00} - T_{11} - T_{22} \\
            T_{01} + T_{10} \\
            T_{02} + T_{20} \\
            T_{11} - T_{22} \\
            T_{12} + T_{21}
        \end{pmatrix}}
    \end{equation*}
    \caption[Decomposing a vector space into irreducible representation subspaces]{Decomposing a vector space into irreducible representation subspaces. The colored matrices represent an arbitrary element of an $\oth$ representation; the vectors show the components in the basis of the matrix representation. Left: An element of $\oth$ acts via the tensor product representation $\tens{R}\kr\tens{R}$, which is reducible. Right: A change of basis matrix $\tens{S}_\rho$ exists that decomposes the tensor product representation into a direct sum of irreducible representations. Some of these irreducible subspaces may be ignored given index permutation symmetries, and particular component values may be fixed by linear tensor constraints. Visualization inspired by \citep{smidt_intuition_2023}.}
    \label{fig:decompose-to-irreps}
\end{figure}

We introduce an example to illustrate the above ideas. Let $\tens{T}$ be a second-order tensor in three dimensions. The nine components of $\tens{T}$ belong to a nine-dimensional vector space $V$. In the Cartesian basis, these components transform under a rotation in $\oth$ as
\begin{equation*}
    T_{ij}' = \sum_{k, l}R_{ik}R_{jl}T_{kl}.
\end{equation*}
To see how this action corresponds to a tensor product representation, note that the right-hand side may be written as $(\tens{R}\kr\tens{R})\vflat\qty(\tens{T})$, where $\kr$ denotes the Kronecker product of two matrices, and $\vflat\qty(\tens{T})$ denotes the C-style flattening of $\tens{T}$ (the serialization of the tensor components such that the last index varies fastest). This matrix-vector product is illustrated in Fig.~\ref{fig:decompose-to-irreps}. The tensor product representation of two irreducible representations is generally not irreducible, as can be seen in the figure by the fact that $\tens{R}\kr\tens{R}$ is a dense matrix.

As $\oth$ is compact, any finite-dimensional representation of $\oth$ can be decomposed as a direct sum of irreducible representations. That is, a change of basis exists such that all elements of the representation assume a block-diagonal form, with each block (an instance of $\hat{\tens{D}}^{(\ell)}(\tens{R})$) operating on an irreducible subspace (a spherical tensor) independently. These irreducible subspaces are sometimes called \textit{steerable} vector spaces to reflect the additional structure supplied by the representation \citep{cohen_steerable_2016,weiler_3d_2018}. In practice, a change of basis matrix can be built from solutions $\tens{S}^{(\ell)}$ that satisfy
\begin{equation}\label{eq:infer-cob}
    \hat{\tens{D}}^{(\ell)}(\tens{R})\tens{S}^{(\ell)} = \tens{S}^{(\ell)}\rho(\tens{R}),\quad\forall\tens{R}\in\oth,
\end{equation}
where $\hat{\tens{D}}^{(\ell)}(\tens{R})$ is an irreducible representation of order-$\ell$ and $\rho(\tens{R})$ is an arbitrary representation, for example $\tens{R}\kr\tens{R}$ (The superscript on $\tens{S}^{(\ell)}$ is added to keep track of which irreducible representation the solution intertwines). According to Schur's lemma, a solution to Eq.~\ref{eq:infer-cob} exists if and only if the irreducible representation of order-$\ell$ is contained in $\rho$. Furthermore, the dimension of the solution space is equal to the number of copies of the irreducible representation of order $\ell$ in the decomposition, known as the multiplicity of the irreducible representation.

A change of basis matrix $\tens{S}_\rho$ for the entire space of tensor components is obtained by solving Eq.~\ref{eq:infer-cob} for increasingly higher $\ell$, stopping when the sum of dimensions of all solutions, including multiplicities, is equal to the dimension of $\rho$. Concretely, $\tens{S}_\rho$ is the square matrix formed by stacking individual solutions vertically:
\begin{equation}\label{eq:stack-cob}
    \tens{S}_\rho \equiv \begin{pmatrix}
        \tens{S}^{(\ell_1)} \\
        \tens{S}^{(\ell_2)} \\
        \vdots \\
        \tens{S}^{(\ell_n)}
    \end{pmatrix}.
\end{equation}
The rows of $\tens{S}_\rho$ that correspond to a given subspace are a basis for that subspace. Choosing the rows to be orthonormal, one can convert the tensor components in $\vflat\qty(\tens{T})$, originally on the Cartesian basis, into the irreducible or spherical basis with the operation $\tens{S}_\rho\vflat\qty(\tens{T})$. The flattened tensors expressed in the spherical basis are thus a direct sum of the spherical tensors. The different bases (Cartesian vs. spherical) and the associated structure of the representations are illustrated in Fig.~\ref{fig:decompose-to-irreps}.

\subsection{Equivariance}\label{sec:EquiVar}

A tensorially consistent RANS model should not depend on the coordinate system used to express the flow variables \citep{spalart_old-fashioned_2023}. For example, a second-order tensor-valued function $\b{f}$ of a second-order tensor $\b{x}$ should satisfy $\tens{R}\b{f}(\b{x})\tens{R}^{\mathrm{T}} = \b{f}(\tens{R}\b{x}\tens{R}^{\mathrm{T}})$ for an arbitrary rotation $\tens{R}$. Intuitively, this says that the change in the component values from evaluating $\b{f}$ on a transformed input should be entirely due to the change in \textit{perspective} stemming from the rotation of the coordinate system, such that the underlying physics implied by the function remains unchanged. For this reason, such functions are often referred to as isotropic tensor functions, which are simply functions that are equivariant under the action of the $\oth$ symmetry group. Isotropic tensor \textit{functions} are not to be confused with isotropic \textit{tensors}, which are tensors whose components are the same in all rotated coordinate systems. The use of the term ``isotropic" across these contexts has historically caused some confusion \citep{lumley_stochastic_2007}.

\begin{define}[Equivariance]
    Let $X$ and $Y$ be vector spaces. The function $\phi:X\to Y$ is equivariant with respect to a group $G$ if it satisfies 
    \begin{equation*}
        \rho^Y(g)\qty[\phi(x)] = \phi\qty(\rho^X(g)[x]),\quad\text{or diagrammatically}\quad
        \begin{tikzcd}
            X \arrow[r, "\phi"] \arrow[d, "{\rho^X(g)}", swap] & Y \arrow[d, "{\rho^Y(g)}"] \\
            X \arrow[r, "\phi"] & Y
        \end{tikzcd}
    \end{equation*}
    for all $g \in G$ and all $x \in X$, where $\rho^X:G\to GL(X)$ and $\rho^Y:G\to GL(Y)$ are representations of $G$ on $X$ and $Y$. Invariance is a special case of equivariance in which $\rho^Y$ is the trivial representation.
\end{define}

The tensor basis method mentioned in Sec.~\ref{sec:tbm-bg} ensures the equivariance of the model by representing tensor functions as linear combinations of Cartesian tensors. Unfortunately, as discussed before, determining the basis tensors can be extremely difficult, with current applications relying on extensive research into the enumeration of bases for second and third-order tensors with particular symmetries \citep{smith_isotropic_1971,pennisi_third_1992,spencer_theory_1958}. ENNs provide an alternative means to enforce equivariance, and as will be shown shortly, this can be done in a way that guarantees additional tensor relationships via hard constraints in the learning framework. 

\subsection{Equivariant neural networks}\label{sec:enns}

\begin{figure}
    \centering
    \includegraphics[width=\textwidth]{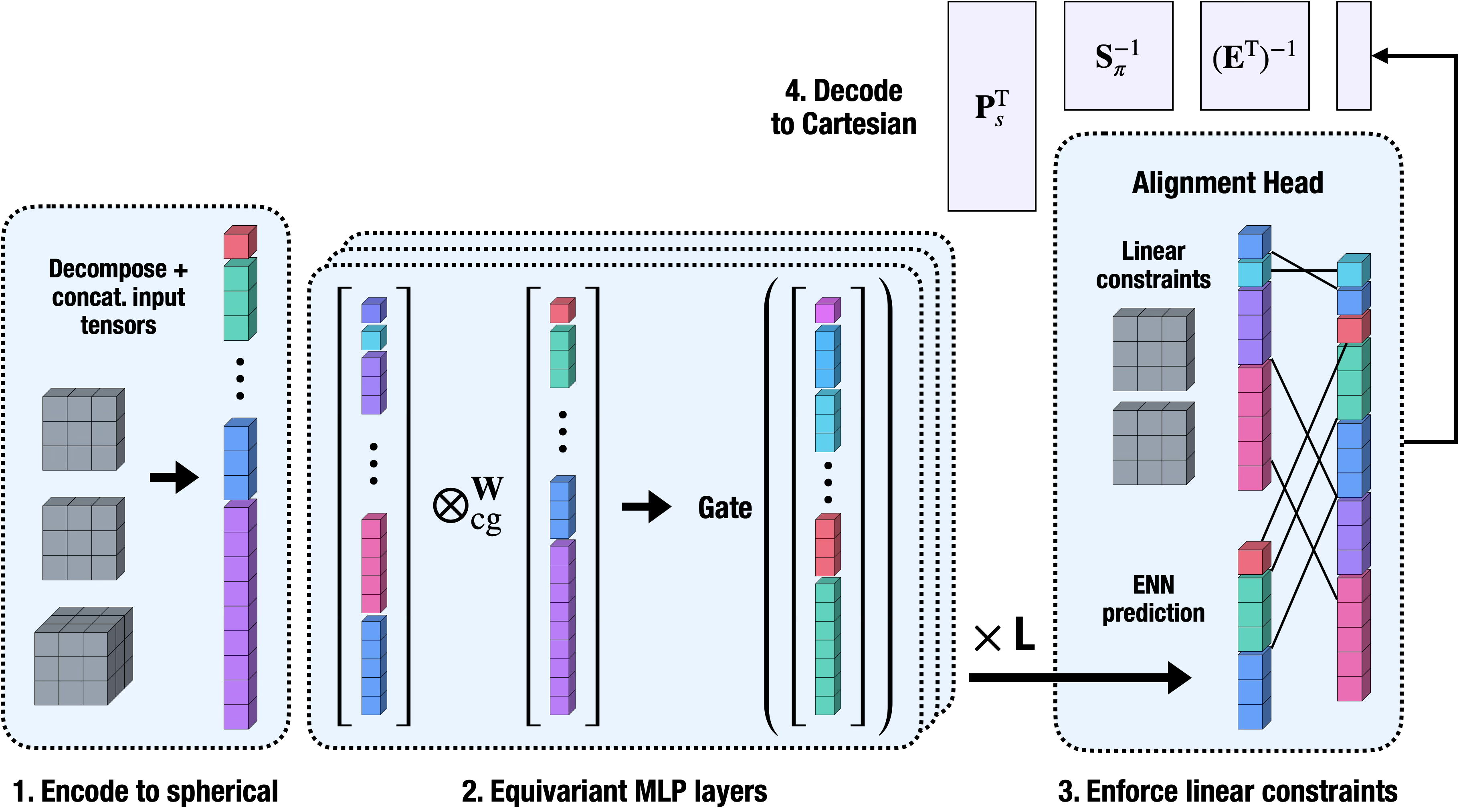}
    \caption{ENN architecture.}
    \label{fig:enn}
\end{figure}
The ENNs considered here use steerable vector spaces to inform the design of equivariant layers. A general recipe for building an equivariant layer $\phi$ is the following:

\begin{enumerate}
    \item Define the input and output spaces of the layer, say $X$ and $Y$, respectively.
    \item Specify the representations $\rho^X$ on $X$ and $\rho^Y$ on $Y$.
    \item Ensure $\phi$ satisfies the equivariance constraint $\rho^Y\circ\phi = \phi\circ\rho^X$.
\end{enumerate}

Step 1 is the same as would be necessary in a traditional neural network, while steps 2 and 3 introduce the additional structure that constrains the network to be equivariant. The irreducible representations of a group represent the fundamental building blocks of all representations, so they can be used to specify $X$, $Y$, $\rho^X$, and $\rho^Y$ in the above framework. It can be shown that irreducible representations of $\oth$ are $(2\ell + 1)$-dimensional and can be indexed by their order $\ell\in \{0, 1, \dots\}$ and their parity $p$, which can be even ($p = 1$) or odd ($p = -1$).

Let the input space of an equivariant layer be
\begin{equation*}
    X = \bigoplus_i V_{\ell_i},
\end{equation*}
where $V_{\ell_i}$ is a steerable vector space of dimension $2\ell_i + 1$. As discussed above, there exists a basis for this space such that elements of $\oth$ act on vectors in the space as
\begin{equation*}
    \hat{\tens{D}}(\tens{R}) = \bigoplus_i\hat{\tens{D}}^{(\ell_i)}(\tens{R}),
\end{equation*}
which is a block-diagonal matrix with blocks of shape $(2\ell_i +1)\times(2\ell_i + 1)$. A vector $\b{x}\in X$ can be viewed as a concatenation of spherical tensors $\b{x}^{(\ell_i)}\in V_{\ell_i}$, each of which has $(2\ell_i + 1)$ components $x^{(\ell_i)}_m$, where $-\ell_i \le m \le \ell_i$ (The parities $p_i$ are suppressed for clarity).

An expressive way to interact two spherical tensors $\b{x}^{(\ell_1)}$ and $\b{y}^{(\ell_2)}$ is via the Clebsch-Gordan tensor product $\cg^w$, which is a bilinear, equivariant operation. It is defined by
\begin{equation}\label{eq:cg-tp}
    (\b{x}^{(\ell_1)}\cg^w\b{y}^{(\ell_2)})_{m_3}^{(\ell_3)} = w\sum_{m_1 = -\ell_1}^{\ell_1}\sum_{m_2 = -\ell_2}^{\ell_2}C^{(\ell_3, m_3)}_{(\ell_1, m_1)(\ell_2, m_2)}x_{m_1}^{(\ell_1)}y_{m_2}^{(\ell_2)},
\end{equation}
where $C^{(\ell_3, m_3)}_{(\ell_1, m_1)(\ell_2, m_2)}$ are the Clebsch-Gordan coefficients and $w$ is a single learnable scalar weight. The Clebsch-Gordan coefficients are only nonzero when the types $(\ell_1, \ell_2, \ell_3)$ satisfy the selection rule $\abs{\ell_1 - \ell_2} \le \ell_3 \le \ell_1 + \ell_2$. For $\oth$, parities combine multiplicatively: $p_3 = p_1 p_2$. Each viable triple $(\ell_1, \ell_2, \ell_3)$ is often called a path \citep{geiger_e3nn_2022}, and Eq.~\ref{eq:cg-tp} is written for a single such path. The input of spherical tensors to the Clebsch-Gordan tensor product can be parts of the same input vector (recall that an input vector is a concatenation of multiple spherical tensors) or two different input vectors. In either case, typically many pairs of spherical tensors can be combined in the product, each pair generally producing several output tensors according to the selection rules.

The output space of the layer is defined similarly to the input space as
\begin{equation*}
    Y = \bigoplus_i V_{\ell_i},
\end{equation*}
where $V_{\ell_i}$ are some subset of steerable vector spaces assigned by the tensor product. That is, the output $(\b{x}^{(\ell_1)}\cg^w\b{y}^{(\ell_2)})^{\ell_3}$ of each path is a steerable vector belonging to some $V_{\ell_i}$, and one may choose which subset of these to concatenate into an output vector $\b{y}\in Y$, typically by fixing a maximum frequency $\ell_{\max}$. In this work, the nonlinear gating mechanism of \citep{brandstetter_geometric_2021} is applied to the output $\b{y}$, and the result is used as input to the next layer. In this way, working with irreducible representations facilitates the design of modular layers suitable for a neural network.

\subsection{Enforcing tensor constraints}\label{sec:constraints}

It is possible to enforce at least three distinct types of tensor constraints in the architecture of an ENN:

\begin{enumerate}
    \item Equivariance
    \item Index permutation symmetries
    \item Linear constraints
\end{enumerate}

The first of these is enforced by constructing the network as a composition of only equivariant functions, for example, the Clebsch-Gordan tensor product and equivariant gate discussed in Sec.~\ref{sec:enns}. Although implied by the name, it is included in the above list to make clear the fact that these are indeed separate types of constraints, in the sense that enforcing one does not automatically enforce the others. Conceptually, the three classes correspond to geometric, structural, and physical modeling constraints, respectively.

\subsubsection{Index permutation symmetries}\label{sec:index-perm}

It is common for certain components of a tensor to always be equal (or related by a minus sign), either by definition or as a consequence of physical symmetry. Such constraints are usually specified with a statement like $T_{ij} = T_{ji}$ or $T_{ij} = -T_{ji}$, which states that a second-order tensor $\tens{T}$ is symmetric or antisymmetric, respectively. The set of symmetry operations on the tensor indices constitutes a permutation group $G$.

Viewing the space of all possible tensor components as a vector space $V$, one can write down a representation $\rho_p$ of $G$ as a set of permutation matrices with dimension equal to the number of tensor components. These matrices describe how the tensor components permute under a symmetry operation on the indices. As with any finite-dimensional representation, $\rho_p$ can be decomposed into irreducible representations. One may also write down a one-dimensional, and hence irreducible, representation $\rho_s$ that stores only the signs associated with each index permutation.

A key result from representation theory is that for any finite group $G$ with representation $\rho$, the operator
\begin{equation*}
    \tens{P}_j = \frac{d_j}{\abs{G}}\sum_{g\in G}\chi_{\rho_j}(g)^*\rho(g),
\end{equation*}
where $*$ denotes complex conjugation, is a projection operator onto the subspace that transforms as the irreducible representation $\rho_j$. Here, $d_j$ and $\chi_{\rho_j}$ are the dimension and character, respectively, of $\rho_j$. The projector $\tens{P}_s$, obtained by taking $\rho_j = \rho_s$ and $\rho = \rho_p$, projects vectors in the space $V$ of all possible tensor components into a subspace $V_s \subseteq V$ with dimension equal to the true number of degrees of freedom in the tensor after removing those fixed by index permutation symmetry. In practice, we realize this projection by choosing a basis for its image, yielding a matrix $\tens{P}_s \in\R^{\operatorname{dim}(V_s)\times\operatorname{dim}(V)}$ that maps a full component vector to reduced coordinates in $V_s$. We choose this basis to be orthonormal, so that $\tens{P}_s\tens{P}_s^{\mathrm{T}} = \tens{I}_{\dim(V_s)}$ and $\tens{P}_s^{\mathrm{T}}$ embeds reduced coordinates back into $V$.

Sec.~\ref{sec:tensors} explained how the vector space $V$, carrying an $\oth$ representation $\rho$, can be decomposed into irreducible subspaces. The same can be done for the subspace $V_s$, which carries a representation $\pi$ containing a subset of the irreducible subspaces in $\rho$. Just as one could solve Eq.~\ref{eq:infer-cob} with $\rho$ to obtain $\tens{S}_\rho$ in Eq.~\ref{eq:stack-cob}, the same can be done using $\pi = \tens{P}_s\rho\tens{P}_s^{\mathrm{T}}$ to obtain $\tens{S}_{\pi}$.

It is important to note that the projector and change-of-basis matrices are tensor-specific: for each tensor $\tens{X}$, $\tens{P}_s$ is constructed from the index-permutation symmetry of $\tens{X}$, and $\tens{S}_{\pi}$ is constructed from the corresponding $\oth$ representation on the reduced component space $V_s$. Hence, we introduce the superscript $[\tens{X}]$ to indicate which tensor these operators act on. The transformation that expresses a flattened tensor $\vflat\qty(\tens{T})$ in the spherical basis, accounting for index permutation symmetries, is therefore $\tens{S}_{\pi}^{[\tens{T}]}\tens{P}_s^{[\tens{T}]}$. One can use an ENN to output components in the subspace $V_s$ and transform back to $V$ to obtain predictions satisfying index permutation symmetries. An implementation of this is available in the \c{CartesianTensor} class of the \c{e3nn} library \citep{geiger_e3nn_2022}.

\subsubsection{Linear tensor constraints}\label{sec:lin-cc}

In this section, a novel algorithm and ENN layer are introduced that can exactly enforce linear tensor constraints on top of equivariance and index permutation symmetries. Linear tensor constraints arise in relationships that involve contractions of one or more pairs of indices of a tensor, for example the condition that a tensor be trace-free. In the language of many deep learning libraries, these could be called ``einsum''-style constraints.

As motivation for the procedure, consider the example of a second-order tensor $\tens{T}$ again. One can see from Fig.~\ref{fig:decompose-to-irreps} that if $\tens{T}$ is symmetric, the components of the $\ell = 1$ subspace are all identically zero, so these three degrees of freedom are fixed. Since this constraint arises due to an index permutation symmetry, it is taken care of by the projection operator $\tens{P}_s$ from Sec.~\ref{sec:index-perm} and does not need to be identified manually. However, the idea of looking for physically meaningful component patterns on the spherical basis underlies the method for enforcing linear constraints.

In particular, note that the component of the irreducible representation $\ell = 0$ is precisely the trace. All one must do to make a trace-free prediction for $\tens{T}$ is to set $T_0^{(0, p)} = 0$ in the spherical basis before transforming back to the Cartesian basis. However, this example represents a convenient special case: since the irreducible representation $\ell = 0$ is one-dimensional, \textit{any} change of basis $\tens{S}_\rho$ in the spherical basis makes this component easily identifiable as the trace. If linear constraints involve higher-dimensional subspaces, in which infinitely many choices are available for the basis (beyond scaling), a more systematic approach is needed to identify the appropriate components to fix in the spherical basis.

To this end, suppose that $\tens{T}$ is an $n$th-order tensor that satisfies one or more linear constraints such as $T_{ijkk} = K_{ij}^{(1)}$ or $\epsilon_{ijk}T_{klm} = K_{ijlm}^{(2)}$ for some Cartesian tensors $\tens{K}^{(1)}$ and $\tens{K}^{(2)}$. Each constraint may be written as $\tgreek{\Lambda}^{(i)}\vflat\qty(\tens{T}) = \vflat\qty(\tens{K}^{(i)})$ for some constant matrix $\tgreek{\Lambda}^{(i)}$, where each row of $\tgreek{\Lambda}^{(i)}$ encodes one component of the relation. For example, if $\tens{T}$ is a second-order symmetric tensor, the row representing a trace-free constraint would be $(1, 0, 0, 0, 1, 0, 0, 0, 1)$. Importantly, if $\tens{T}$ is a tensor function, the prescribed values $\tens{K}^{(i)}$ may depend on the input.

Since each $\tens{K}^{(i)}$ is a Cartesian tensor, with its own index permutation symmetries, its components may be decomposed via left multiplication by $\tens{S}_{\pi}^{[\tens{K}^{(i)}]}\tens{P}_s^{[\tens{K}^{(i)}]}$, where we have applied the tensor-specific notation introduced in Sec.~\ref{sec:index-perm}. The constraint equation can be written as 
\begin{equation}\label{eq:single-tc}
    \hat{\tgreek{\Lambda}}^{(i)}\vflat\qty(\tens{T}) = \tens{S}_{\pi}^{[\tens{K}^{(i)}]}\tens{P}_s^{[\tens{K}^{(i)}]}\vflat\qty(\tens{K}^{(i)}),
\end{equation}
where $\hat{\tgreek{\Lambda}}^{(i)} = \tens{S}_{\pi}^{[\tens{K}^{(i)}]}\tens{P}_s^{[\tens{K}^{(i)}]}\tgreek{\Lambda}^{(i)}$. All the tensor constraints can be collected into a single statement 
\begin{equation*}
    \hat{\tgreek{\Lambda}}\vflat\qty(\tens{T}) = \b{k}
\end{equation*}
by stacking the rows of $\hat{\tgreek{\Lambda}}^{(i)}$ and the right-hand side of Eq.~\ref{eq:single-tc} for all $i$, specifically
\begin{equation*}
    \hat{\tgreek{\Lambda}} \equiv \begin{pmatrix}
        \hat{\tgreek{\Lambda}}^{(1)} \\
        \hat{\tgreek{\Lambda}}^{(2)} \\
        \vdots
    \end{pmatrix},\qquad \b{k} \equiv \begin{pmatrix}
        \tens{S}_{\pi}^{[\tens{K}^{(1)}]}\tens{P}_s^{[\tens{K}^{(1)}]}\vflat\qty(\tens{K}^{(1)}) \\
        \tens{S}_{\pi}^{[\tens{K}^{(2)}]}\tens{P}_s^{[\tens{K}^{(2)}]}\vflat\qty(\tens{K}^{(2)}) \\
        \vdots
    \end{pmatrix}.
\end{equation*}

The key observation is that $\tens{T}$ and each $\tens{K}^{(i)}$ can be decomposed into irreducible representations, so there exists a particular choice of spherical bases in which the linear constraints correspond to fixing a subset of spherical components of $\tens{T}$ to prescribed values (those contained in $\b{k}$). What is needed is to find a common set of bases for the subspaces produced by 1) decomposing $\tens{T}$ into irreducible representations and 2) decomposing the tensors generated via contraction, e.g. $T_{ijkk}$ or $\epsilon_{ijk}T_{klm}$, into irreducible representations. Then it becomes clear which spherical components of $\tens{T}$ are fixed by linear constraints and which degrees of freedom remain to be predicted.

Alg.~\ref{alg:lin-ch} specifies a procedure for computing a matrix $\tens{E}$ that performs this basis alignment. Specifically, it returns $\tens{E}$, $\bm{\alpha}$, and $\bm{\beta}$ such that the components of
\begin{equation*}
    \tilde{\tens{T}} = \tens{E}^{\mathrm{T}}\tens{S}_{\pi}^{[\tens{T}]}\tens{P}_s^{[\tens{T}]}\vflat\qty(\tens{T})
\end{equation*}
at indices $\bm{\alpha}$ equal the components of $\b{k}$ at indices $\bm{\beta}$, that is, $\tilde{\tens{T}}[\bm{\alpha}] = \b{k}[\bm{\beta}]$. In practice, one predicts the components of $\tilde{\tens{T}}$ at the indices not in $\bm{\alpha}$, fills in the remaining components by setting $\tilde{\tens{T}}[\bm{\alpha}] = \b{k}[\bm{\beta}]$, and applies the inverse transformation $\qty(\tens{P}_s^{[\tens{T}]})^{\mathrm{T}}\qty(\tens{S}_{\pi}^{[\tens{T}]})^{-1}(\tens{E}^{\mathrm{T}})^{-1}\tilde{\tens{T}}$ to produce a tensor in the Cartesian basis that satisfies the constraints, as shown in stage 3 of Fig.~\ref{fig:enn}.

\begin{algorithm}
    \LinesNumbered
    \SetKwInOut{KwInput}{Input}
    \SetKwInOut{KwOutput}{Output}
    \ResetInOut{Output}
    \KwInput{Lists $\{\ell_i\}$, $\{\tens{S}_\pi^{(\ell_i)}\}$ and projector $\tens{P}_s^{[\tens{T}]}$ for tensor $\tens{T}$; Constraint matrix $\hat{\tgreek{\Lambda}}$}
    \KwOutput{Alignment matrix $\tens{E}$; Lists $\bm{\alpha}, \bm{\beta}$ of alignment indices}

    \DontPrintSemicolon
    \SetInd{0.5em}{1em}
    \SetNlSkip{0.5em}

    \BlankLine
    \emph{Computes a change of basis matrix between the spherical basis of $\tens{T}$ and the basis implied by the linear constraint matrix $\hat{\tgreek{\Lambda}}$.}\;
    \BlankLine
    \SetKwComment{Comment}{// }{}
    \SetCommentSty{small}

    \Begin{
        Initialize $\tens{E} \gets \tens{I}$\Comment*[r]{Identity matrix with dimension of $V_s$}
        Initialize $\bm{\alpha} \gets [\ ], \bm{\beta} \gets [\ ]$\Comment*[r]{Empty lists of alignment indices}
        Initialize $j \gets 0$\Comment*[r]{Used in for loop to build $\bm{\alpha}$}
        $\tens{B} \gets \tens{P}_s^{[\tens{T}]}\hat{\tgreek{\Lambda}}^{\mathrm{T}}$\Comment*[r]{Cols.\ of $\tens{B}$ are RHS vectors for Line 11}
        \For{$\ell$ \textnormal{such that} $\exists \tens{S}^{(\ell)}\in \{\tens{S}^{(\ell_i)}\}$}{
            $m \gets \operatorname{Count}\qty(\ell\in\{\ell_i\})$\Comment*[r]{Count multiplicity of $\ell$ in list $\{\ell_i\}$}
            $d \gets 2\ell + 1$\Comment*[r]{Dimension of irreducible rep. at order $\ell$}
            $\tens{A} \gets \operatorname{Stack}\qty(\tens{S}^{(\ell_i)})$\Comment*[r]{Stack all $\tens{S}^{(\ell_i)}$ for which $\ell_i = \ell$}
            $\tens{X}, \b{r} \gets \operatorname{LeastSquares}\qty(\tens{A}^{\mathrm{T}}, \tens{B})$\Comment*[r]{Store solutions in $\tens{X}$, residuals in $\b{r}$}
            \If(\tcp*[f]{Count solutions using tolerance $\tau$}){$(\operatorname{Count}\qty(\b{r} < \tau) = md)$}{
                $\tens{E}[j:j + md, j:j + md] = \tens{X}[:, \operatorname{Where}\qty(\b{r} < \tau)]$\;
                $\bm{\alpha}.\operatorname{AddElements}\qty(j: j + md)$\;
                $\bm{\beta}.\operatorname{AddElements}\qty(\operatorname{Where}\qty(\b{r} < \tau))$
            }
            $j \gets j + md$\;
        }
        \Return{$\tens{E}, \bm{\alpha}, \bm{\beta}$}\;
    }
    \caption{Linear constraint alignment matrix}
    \label{alg:lin-ch}
\end{algorithm}

The following details accompany the indicated line numbers from Alg.~\ref{alg:lin-ch}:
\begin{itemize}
    \item Input: The index permutation symmetries of $\tens{T}$ determine $\tens{P}_s$, which is used to construct the representation $\pi = \tens{P}_s\rho\tens{P}_s^{\mathrm{T}}$ that acts on $V_s$. Eq.~\ref{eq:infer-cob} is solved for increasing values of $\ell$ to find a list of solutions $\{\tens{S}_\pi^{(\ell_i)}\}$ with orders $\{\ell_i\}$ until the cumulative dimension of the solutions equals the dimension of $V_s$. Importantly, the rows of $\hat{\tgreek{\Lambda}}$ should be made linearly independent by pruning redundant constraints so that $\tens{E}$ is invertible (see App.~\ref{sec:inv-E}).
    \item Line 6: The number of columns of $\hat{\tgreek{\Lambda}}$ is equal to the number of components of $\tens{T}$ (Each row represents a constraint and is specified with respect to $V$, the space of all tensor components). To express the rows with respect to $V_s$, and thereby account for index permutation symmetries, the projector $\tens{P}_s$ must act on the columns of $\hat{\tgreek{\Lambda}}$, hence the transposition.
    \item Lines 11: $\operatorname{LeastSquares}\qty(\tens{A}^{\mathrm{T}}, \tens{B})$ denotes a batched least squares problem, in which a solution is sought for each column of $\tens{B}$ independently.
    \item Lines 12-16: If the least squares residual is zero (in practice, less than a small tolerance such as $\tau = 10^{-10}$) for a group of columns, the spherical bases at order $\ell$ span a space in which components of the linear constraints live. For each $\ell$ block, either no matching columns are found, or exactly $md$ are found. The solutions corresponding to zero-residual columns relate the spherical and constraint bases, providing the desired mapping needed to equate components.
\end{itemize}

Formally, Alg.~\ref{alg:lin-ch} accounts for the fact that the decomposition of an isotypic component of a group representation is not unique. An isotypic component of a representation is the direct sum of all irreducible subspaces transforming according to a specific irreducible representation $\ell$, or intuitively the number of copies of a particular irreducible representation that appear within a larger representation. If the multiplicity of an order-$\ell$ irreducible representation is $k$, the representation matrices on this isotypic component can be written as $\tens{I}_k \kr \hat{\tens{D}}^{(\ell)}(\tens{R})$. Schur's lemma requires that a change of basis in this subspace must be of the form $\tens{U} = \tens{U}_k \kr \tens{I}_{2\ell + 1}$, where $\tens{U}_k$ is any $k\times k$ unitary matrix, showing that there is a $U(k)$ freedom, where $U(k)$ is the unitary group of degree $k$, to ``mix" the $k$ copies of the order-$\ell$ irreducible representations amongst themselves (If $\tens{U}$ does not map between orthonormal bases, $\tens{U}_k$ can belong to a larger group of matrices). Alg.~\ref{alg:lin-ch} determines whether two isotypic components are expressed differently under this freedom and provides a map to align them.

\subsection{Architecture}\label{sec:architecture}

An overview of the complete ENN architecture used here is shown in Fig.~\ref{fig:enn}. In stage 1 of the network, Cartesian tensors are decomposed via left multiplication by the matrix $\tens{S}_{\pi}^{[\tens{X}]}\tens{P}_s^{[\tens{X}]}$ appropriate to each input tensor $\tens{X}$. Stage 2 contains a sequence of $L$ equivariant multi-layer perceptron (MLP) layers, where $\cg^\tens{W}$ denotes the fully connected Clebsch-Gordan tensor product of the hidden representation and the original input. Fully connected refers to the terminology used to describe the \c{FullyConnectedTensorProduct} class of the \c{e3nn} library \citep{geiger_e3nn_2022}. The bold $\tens{W}$ denotes the collection of weights for all paths, in contrast to the scalar $w$ in Eq.~\ref{eq:cg-tp}, which is written for a single path. Intermediate layers consist of even and odd parity irreducible representations for orders up to and including $\ell_{\max} = 6$, with multiplicities chosen such that each order has roughly the same number of components \citep{brandstetter_geometric_2021}.

The different types of tensor constraints described in Sec.~\ref{sec:constraints} are enforced via single matrix multiplications at the encoding and decoding stages of the network and hence represent lightweight modifications relative to the processing block. Additionally, the matrices involved need to be computed only once offline. Stage 3 of the figure shows the novel output layer, in which predictions from the ENN are concatenated with constraint-determined component values in a common spherical basis before being decoded in stage 4 back into the Cartesian space.

The architecture is implemented with PyTorch using the \c{e3nn} library \citep{geiger_e3nn_2022}. As the new layer operates on the output of the MLP blocks, it has the advantage that no code changes to the library are required.

\section{Results}\label{sec:results}

In this section, the ENN tensor modeling formulation described in Sec. \ref{sec:methods} is applied to RANS closure modeling in the limit of RDT, using the framework advanced by Kassinos and Reynolds \citep{kassinos_one-point_2001,kassinos_structure-based_1995}. Across the three closure modeling problems required in this setting ($\tens{M}$, $\tens{L}$, and $\tens{J}$), there are 405 total tensor components that must be predicted. Exploiting the strategies in Sec.~\ref{sec:constraints}, only 47 of these need to be predicted independently. The constraints governing the remaining 88\% are enforced exactly, regardless of training.

\subsection{Modeling objective}\label{sec:modeling-obj}

Eq.~\ref{eq:rdt-q} requires closure models for 
\begin{equation}\label{eq:required-closures}
    \tens{M}\qty(\rst, \tens{D}, \tens{Q}^*),\quad \tens{L}\qty(\tens{D}, \tens{Q}^*),\qq{and}\tens{J}\qty(\rst, \tens{D}, \tens{Q}^*).
\end{equation}
Several considerations help frame the problem:
\begin{enumerate}
    \item The learned models implicitly form higher-order polynomials of the tensors. To ensure that these models are consistent with dimensional analysis, all tensors in Eq.~\ref{eq:required-closures} are normalized by $q^2$.
    \item All quantities shown are (generalized) Cartesian tensors, so their components must satisfy the implied transformation rules, namely Eq.~\ref{eq:ct-def} on the Cartesian basis or Eq.~\ref{eq:sph-def} on the spherical basis. This is the core principle that motivates the use of ENNs.
    \item Any other structural constraints should also be respected. That is, the outputs should satisfy the index permutation symmetries listed in table~\ref{tab:tensors}.
    \item Known physical constraints should be satisfied. The relevant constraints for this case are shown in table~\ref{tab:constraints}.
    \item The functions in Eq.~\ref{eq:required-closures} are local in the sense that they relate quantities at a single position and time. In an inhomogeneous flow, one might consider a dependence on values in a neighborhood instead.
\end{enumerate}

\begin{table}
    \centering
    \begin{minipage}[c]{0.55\textwidth}
        \centering
        \captionsetup{width=\linewidth}
        \begin{tabular}{ccc}
            \toprule
            & Order & Index permutation symmetry \\
            \midrule
            $\tens{M}$ & 4 & $M_{ijpq} = M_{ijqp} = M_{jipq} = M_{jiqp}$ \\
            $\tens{M}^*$ & 4 & Fully symmetric \\
            $\tens{L}$ & 4 & Fully symmetric \\
            $\rst$ & 2 & Fully symmetric \\
            $\tens{D}$ & 2 & Fully symmetric \\
            $\tens{Q}$ & 3 & No symmetry \\
            $\tens{Q}^*$ & 3 & Fully symmetric \\
            \multirow{2}{*}{$\tens{J}$} & \multirow{2}{*}{5} & $J_{ijrpq} = J_{ijrqp} = J_{ijprq}$ \\
            & & $ = J_{ijpqr} = J_{ijqrp} = J_{ijqpr}$ \\
            \bottomrule
        \end{tabular}
        \caption{Tensor symmetries \citep{kassinos_one-point_2001}.}
        \label{tab:tensors}
    \end{minipage}\hfill
    \begin{minipage}[c]{0.45\textwidth}
        \centering
        \captionsetup{width=\linewidth}
        \begin{tabular}{cr@{${}={}$}l}
            \toprule
            & \multicolumn{2}{c}{Constraints} \\
            \midrule
            \multirow{4}{*}{$\tens{M}$} & $M_{ikkj}$ & $0$ \\
            & $M_{ijkk}$ & $\rs_{ij}$ \\
            & $M_{kkij}$ & $D_{ij}$ \\
            & $\sym(\epsilon_{ipq}M_{jqpk})$ & $Q^*_{ijk}$ \\
            \midrule
            $\tens{M}^*$ & $M_{ijkk}$ & $1/6\qty(\rs_{ij} + D_{ij})$ \\
            \midrule
            $\tens{L}$ & $L_{ijkk}$ & $D_{ij}$ \\
            \midrule
            \multirow{6}{*}{$\tens{J}$} & $J_{kkijr}$ & $0$ \\
            & $J_{ikjrk}$ & $0$ \\
            & $J_{kijrk}$ & $0$ \\
            & $\epsilon_{jsr}J_{sirkk}$ & $\rs_{ij}$ \\
            & $\epsilon_{irs}J_{srjkk}$ & $D_{ij}$ \\
            & $\sym(J_{ijrkk})$ & $Q^*_{ijr}$ \\
            \bottomrule
        \end{tabular}
        \caption{Modeling constraints \citep{kassinos_one-point_2001}.}
        \label{tab:constraints}
    \end{minipage}
\end{table}

The considerations 2-4 illustrate why each of the three properties discussed in Sect.~\ref{sec:constraints} is relevant to RANS modeling. Importantly, all of the constraints in tables~\ref{tab:tensors} and \ref{tab:constraints} are enforced exactly by the network architecture and so are obeyed regardless of training, in contrast to the soft constraint strategies mentioned in Sec.~\ref{sec:prior-constraints}.

\subsection{Data generation}\label{sec:data}

Since the target models arise in the evolution equation for $\tens{Q}$ in the rapid distortion limit, the data to train and test the model is obtained from the solution of the RDT equations. The data consists of $N$ data points and targets taken from RDT solutions obtained for a range of velocity gradient tensors at different times during the evolution.

The $k$th data point consists of an input tuple $(\rst_k, \tens{D}_k, \tens{Q}^*_k)$, and the corresponding target is one of $\tens{M}_k$, $\tens{L}_k$, or $\tens{J}_k$ depending on the closure sought. Once this data set is obtained, the training procedure follows a standard supervised learning design. The loss function used here is the mean-squared error between the ENN predictions and the targets, which can be measured on either a spherical basis or a Cartesian basis. 

For homogeneous turbulence, the target tensors $\tens{M}$, $\tens{L}$, and $\tens{J}$ are expressed in terms of the velocity spectrum tensor $\tgreek{\Phi}$ in Eqs.~\ref{eq:M-def} and \ref{eq:LJ-def}. The inputs $\rst$, $\tens{D}$ and $\tens{Q}^*$ can similarly be expressed in terms of $\tgreek{\Phi}$:
\begin{align}
  \rs_{ij} &=\int\Phi_{ij}\,\dd\bm{\kappa}\label{eq:input-def1},\\
  D_{ij}&= \int\Phi_{nn}\frac{\kappa_i\kappa_j}{\kappa^2}\,\dd\bm{\kappa},\\
  Q_{ijk}& = \epsilon_{ipq}\int\Phi_{jq}\frac{\kappa_{p}\kappa_{k}}{\kappa^{2}}\,\dd{\bm{\kappa}},\qquad \tens{Q}^* \equiv \sym(\tens{Q}).\label{eq:input-def3} 
\end{align}
Using these relations, all of the necessary data can be generated from the velocity spectrum tensor $\tgreek{\Phi}$ by integrating over wavevector space.

\begin{figure}
    \centering
    \includegraphics[width=\textwidth]{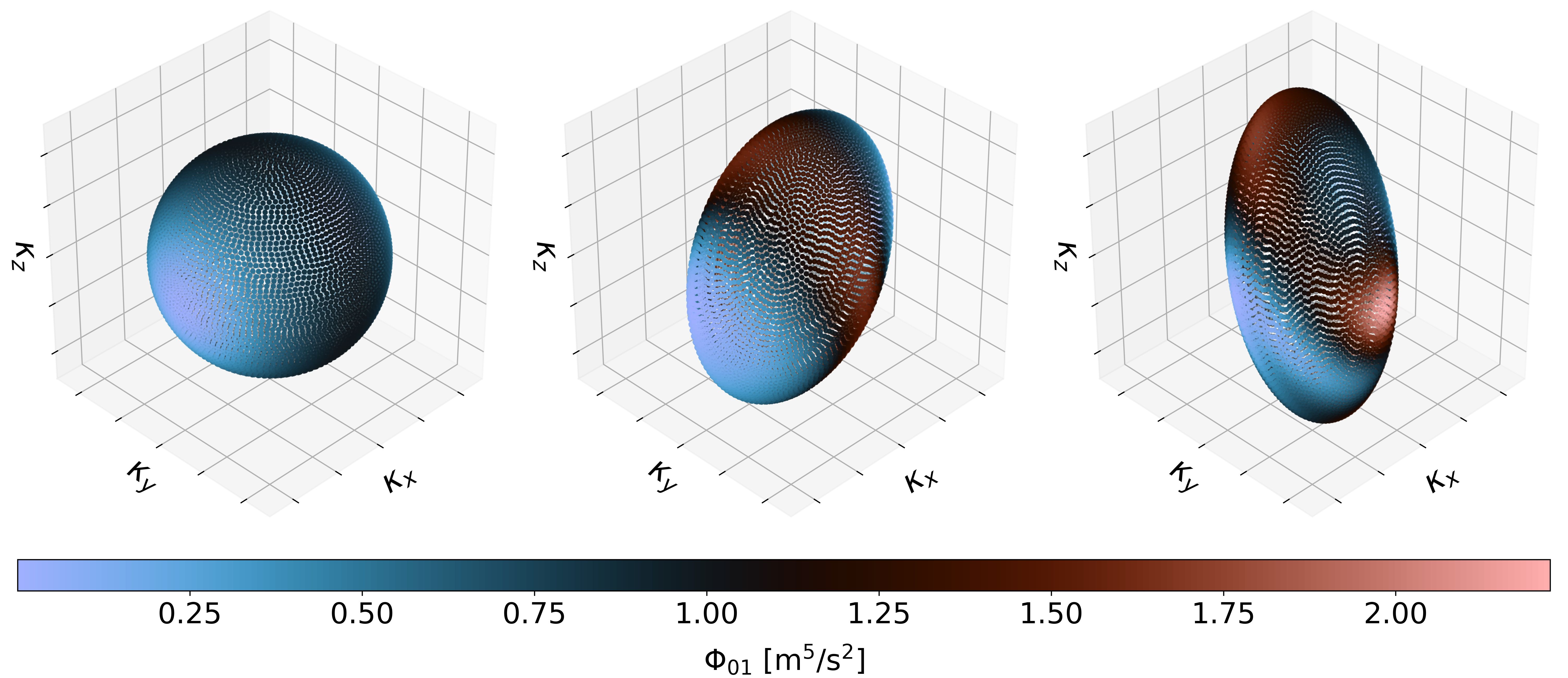}
    \caption{Evolution of wavevectors $\bm{\kappa}(t)$ at which the velocity spectrum tensor $\tgreek{\Phi}$ can be integrated independently. Each marker corresponds to a wavevector, initially located at a spherical design quadrature point on the unit sphere. As time proceeds (left to right in the figure), the wavevectors distort, and the values of $\tgreek{\Phi}$ following the wavevectors evolve.}
    \label{fig:wavevectors}
\end{figure}

In RDT, when the velocity spectrum tensor is evaluated at a wavevector $\bm{\kappa}(t)$, where the wavevector evolves according to
\begin{equation}
    \dv{\kappa_\ell}{t} = -\kappa_j\pdv{U_j}{x_\ell},\label{eq:kappa-ev}
\end{equation}
the spectrum tensor $\Phi_{ij}(\bm{\kappa}(t),t)$ satisfies the following ODE:
\begin{equation}\label{eq:phi-ev}
    \dv{\Phi_{ij}}{t} = -\pdv{U_i}{x_k}\Phi_{kj} - \pdv{U_j}{x_k}\Phi_{ik} + 2\pdv{U_\ell}{x_k}\qty(\frac{\kappa_i\kappa_\ell}{\kappa^2}\Phi_{kj} + \frac{\kappa_j\kappa_\ell}{\kappa^2}\Phi_{ik}).
\end{equation}
Furthermore, because the coefficients in Eq.~\ref{eq:phi-ev} do not depend on the magnitude of $\bm{\kappa}$, the solution for $\tgreek{\Phi}$ with an initial wavevector $\bm{\kappa}_0$ can be used to determine the solution for any initial wavevector $\alpha\bm{\kappa}_0$, provided the initial conditions $\tgreek{\Phi}(\alpha\bm{\kappa}_0,t=0)=\beta\tgreek{\Phi}(\bm{\kappa}_0,t=0)$ for some scalar $\beta$ that depends on $\alpha$. This is true for the isotropic initial conditions used to generate the data, so it is sufficient to solve Eqs.~\ref{eq:kappa-ev} and \ref{eq:phi-ev} for $\bm{\kappa}_0$ restricted to a sphere.     
The initial shell of wavevectors distorts with time (see Fig.~\ref{fig:wavevectors}), but a change of variables allows integrals over the surface to be mapped back to the initial sphere. As long as the initial condition for $\tgreek{\Phi}$ satisfies the condition above, the integrals in Eqs.~\ref{eq:M-def}, \ref{eq:LJ-def}, and \ref{eq:input-def1}--\ref{eq:input-def3} can all be evaluated as weighted integrals over the initial spherical shell.

Data have been generated for 320 different mean velocity gradients, sampled using Sobol sequencing \citep{sobol_distribution_1967}, by integrating Eq.~\ref{eq:phi-ev} for initial wavevectors distributed over the unit sphere at 6~051 spherical design points, which are equal-weight numerical quadrature points that guarantee exact integration on the unit sphere for spherical polynomials up to a certain degree (here, $\le 109$) \citep{womersley_efficient_2017}. Storing the $\tgreek{\Phi}$ solution at 100 time steps per simulation, this procedure produces 32~000 samples for each closure problem.

\subsection{ENN training and evaluation}\label{sec:enn-train-eval}

First, results are presented for the modeling of $\tens{M}^*\qty(\rst, \tens{D})$ and $\tens{M}\qty(\rst, \tens{D}, \tens{Q}^*)$, where $\tens{M}^*$ is the fully symmetric part of $\tens{M}$. The purpose of modeling $\tens{M}^*$ with and without the $\tens{Q}^*$ dependence is to assess the importance of nonlinear dependence on $\tens{Q}^*$ to the model. Note that when $\tens{Q}^*$ is included as a dependence, modeling $\tens{M}$ and $\tens{M}^*$ are equivalent (see Sec.~\ref{sec:modeling-M}). Analogously, results are shown for models of $\tens{L}(\tens{D})$ and $\tens{L}(\tens{D}, \tens{Q}^*)$, and $\tens{J}(\tens{R}, \tens{D})$ and $\tens{J}(\tens{R}, \tens{D}, \tens{Q}^*)$. Further, the $\tens{J}$ model is used to illustrate the role of the $\tens{E}$ matrix produced by Alg.~\ref{alg:lin-ch}.

The \c{e3nn} shorthand ``$m_1\times\ell_1p_1 + m_2\times\ell_2p_2 + \cdots$" is used in this section to describe spherical tensors, where $m_i$, $\ell_i$, and $p_i$ denote the multiplicity, order, and parity, respectively, of the irreducible representations that make up a tensor. For example, $3\times 0\rm{e} + 2\rm{e}$ denotes a spherical tensor that is the direct sum of three irreducible representations $\ell = 0$ and one irreducible representation $\ell = 2$, all of even parity.

As mentioned in Sec.~\ref{sec:data}, the complete data set consists of 32~000 samples from 320 independent simulations. All models are trained on 40\% of this data, chosen by sampling 128 simulations and including all 100 time steps from each. Similarly, the validation data set consists of all the time steps from 32 simulations, or 10\% of the total data. Although viewing the 32~000 samples as a single pool would lead to more diverse training data, splitting based on simulations is preferred because it guarantees that the validation and test sets consist of velocity gradients that have not been seen during training. Figs.~\ref{fig:m}, \ref{fig:l}, and \ref{fig:j} show results for all 320 simulations, spanning the training, validation, and test data. The errors are those at the final time steps of the simulations, chosen because these times represent the greatest deviation from isotropy.

The decision to use a relatively small fraction of the data for training was made to keep training times low while using a small batch size (here 32). The batch size was observed to be one of the most critical hyperparameters, with lower batch sizes leading to better performance. Additional details regarding training and hyperparameters can be found in App.~\ref{sec:train-details}.

\subsubsection[Modeling M]{Modeling $\tens{M}$}\label{sec:modeling-M}

It is shown by \citet{kassinos_structure-based_1995} that $\tens{M}$ satisfies the exact decomposition
\begin{equation}\label{eq:m-decomp}
    \begin{aligned}
        M_{ijkl} &= M_{ijkl}^* \\
        &+ \frac{1}{2}(\epsilon_{zkj}Q_{zil}^* - \epsilon_{zil}Q_{zkj}^*) \\
        &+ \frac{1}{6}\left[(\delta_{il}\delta_{jk} + \delta_{ik}\delta_{lj} - 2\delta_{ik}\delta_{jl})q^2 + 3(\delta_{kl}\rs_{ij} + \delta_{ij}D_{kl}) + \delta_{kl}D_{ij} + \delta_{ij}\rs_{kl}\right. \\
        &- \left.\delta_{il}(\rs_{kj} + D_{kj}) - \delta_{kj}(\rs_{il} + D_{il}) - \delta_{ik}(\rs_{lj} + D_{lj}) - \delta_{lj}(\rs_{ki} + D_{ki})\right],
    \end{aligned}
\end{equation}
so one can either model $\tens{M}$ directly or model $\tens{M}^*$ and use this decomposition to construct $\tens{M}$. In the most general case, either model would include all of $\rst$, $\tens{D}$, and $\tens{Q}^*$ as dependencies.

With the symmetries shown in table~\ref{tab:tensors}, $\tens{M}$ decomposes into a $2\times 0\rm{e} + 1\rm{e} + 3\times 2\rm{e} + 3\rm{e} + 4\rm{e}$ spherical tensor with 36 independent components. The constraints in table~\ref{tab:constraints} fix the components for $\ell\in\{0, 1, 2, 3\}$, leaving only the nine $\ell = 4$ components to be predicted. The $\tens{M}^*$ tensor decomposes into $0\rm{e} + 2\rm{e} + 4\rm{e}$, a spherical tensor with 15 components. Of these, the components of the $\ell = 0$ and $\ell = 2$ irreducible representations are fixed. It is shown in App.~\ref{sec:ms-constraints} that the relations in table~\ref{tab:constraints} for $\tens{M}$, which correspond to 27 independent constraints, translate into six independent constraints for $\tens{M}^*$.

As mentioned in Sec.~\ref{sec:data}, the training procedure for all closure models follows a supervised learning design, and the predicted and target tensors may be compared in either the spherical or Cartesian bases. The former requires less computation and allows us to investigate the effect of enforcing the modeling constraints in table~\ref{tab:constraints} during training. That is, while all constraints should undoubtedly be enforced during inference to guarantee physically valid predictions, this does not necessarily mean that they should be enforced during training. In this discussion and in the remainder of the paper, statements about training ``with and/or without constraints" refers only to enforcement of the \textit{modeling} constraints in table~\ref{tab:constraints}. The geometric (equivariance) and structural (index permutation symmetry) constraints are always enforced.

\begin{figure}
    \centering
    \includegraphics[width=0.49\textwidth]{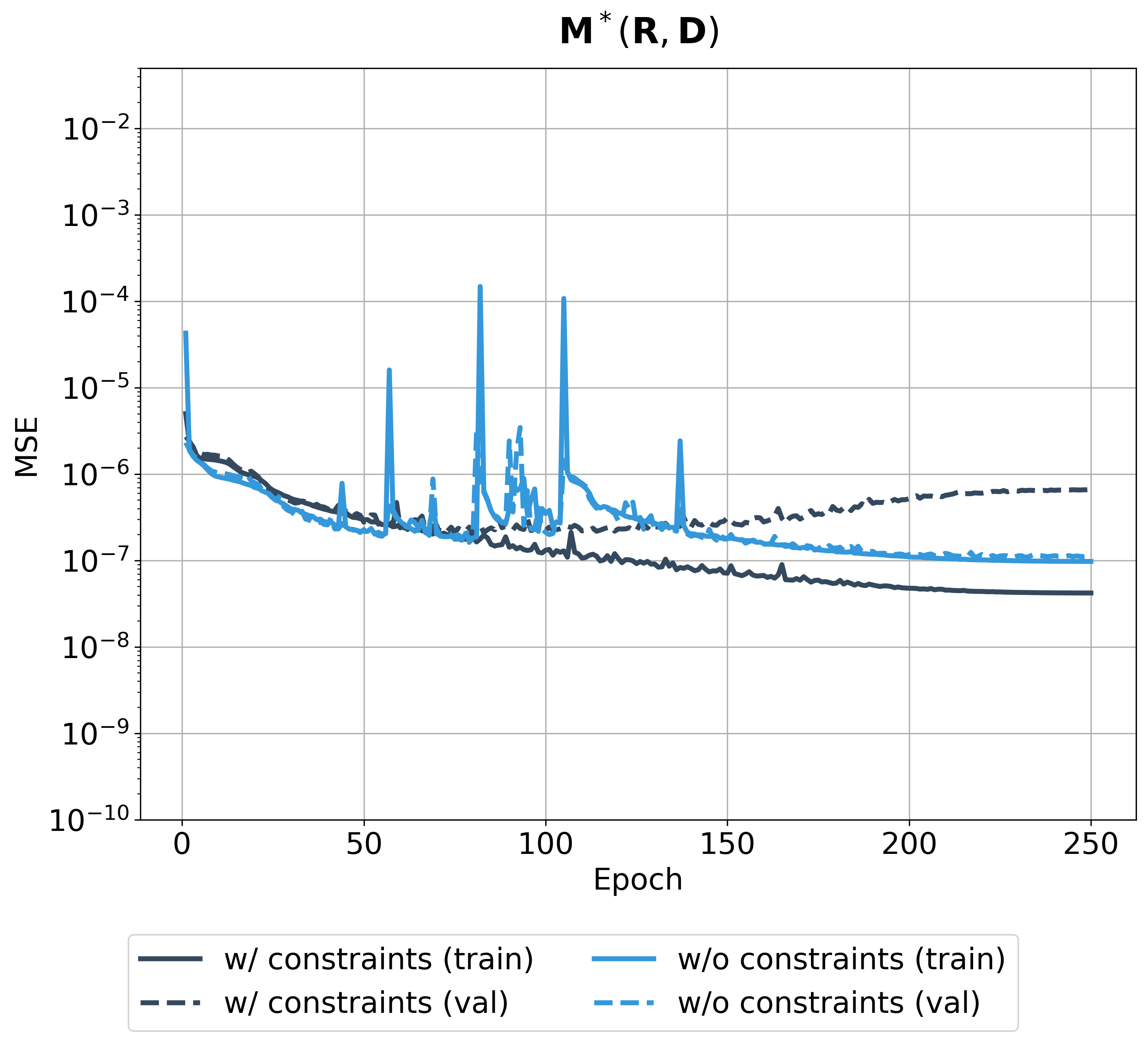}
    \includegraphics[width=0.49\textwidth]{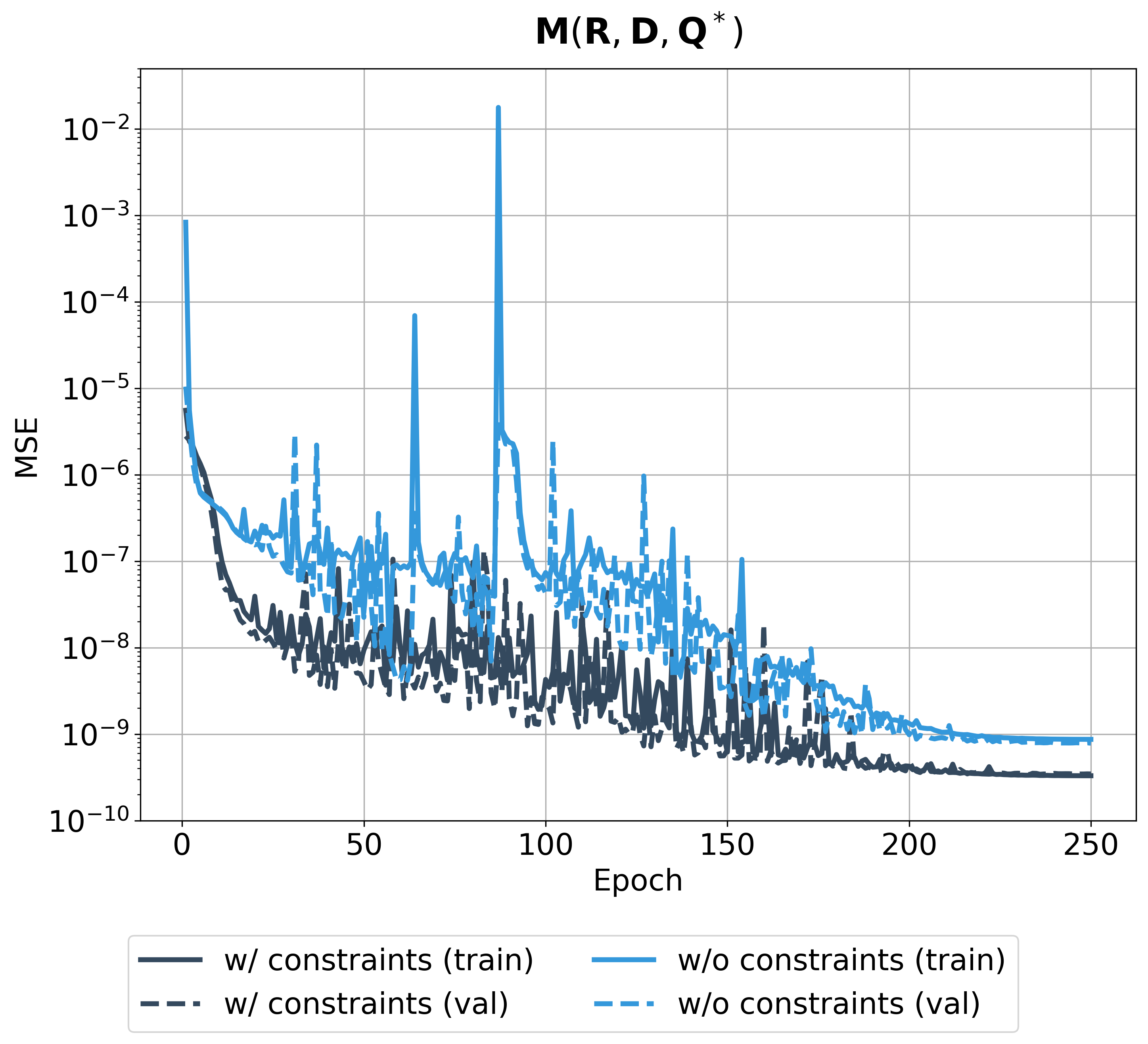}
    \caption{ENN training and validation loss when learning $\tens{M}^*$ (left) and $\tens{M}$ (right) with and without the constraints in table~\ref{tab:constraints} enforced. The ``w/ constraints'' designation refers to learning with supervision from only the nine components of the $\ell = 4$ irreducible representation, while ``w/o constraints" refers to supervising on all 15 components of the $0\rm{e} + 2\rm{e} + 4\rm{e}$ spherical tensor for $\tens{M}^*$, or all 36 components of the $2\times 0\rm{e} + 1\rm{e} + 3\times 2\rm{e} + 3\rm{e} + 4\rm{e}$ spherical tensor for $\tens{M}$.}
    \label{fig:ms_rd_train}
\end{figure}

On the one hand, learning with enforcement requires the network to predict fewer tensor components, potentially making the learning problem easier. For example, when predicting either $\tens{M}^*$ or $\tens{M}$, the only unconstrained components are those of the $\ell = 4$ irreducible representation. Learning this contribution requires an output of nine components, rather than 36 or 15 for learning the full $\tens{M}$ or $\tens{M}^*$ tensors, respectively.
 
On the other hand, if the ENN predicts only the unconstrained components, the gradient signal will come entirely from the highest frequency part of the tensor, which may make the network more prone to find spurious features relevant only to the $4\rm{e}$ components. In this case, learning to predict the components of all irreducible representations may serve as a source of implicit regularization. Since the predictions for each order $\ell$ rely on the same shared internal representation up until the final network layer, the learned internal features must be versatile enough to benefit all tasks, discouraging learning pathways that overfit to the $4\rm{e}$ components.

In practice, it was observed that differences in performance and overfitting behavior when training with or without constraints were dependent on whether $\tens{Q}^*$ was included as a model input. When including $\tens{Q}^*$, no overfitting was observed in a training sweep over 12 model hyperparameter configurations, regardless of whether constraints were enforced. Additionally, training with constraints led to a clear performance benefit. Conversely, when $\tens{Q}^*$ dependence was excluded, overfitting behavior was sometimes observed, both with and without constraint enforcement, and no performance benefit was gained from training with constraints (See Fig.~\ref{fig:ms_rd_train}). In fact, without $\tens{Q}^*$, omitting constraints led to slightly better performance.

One might view these observations, particularly that models with $\tens{Q}^*$ dependence exhibit less overfitting and improved performance from constrained learning, as evidence that $\tens{R}$, $\tens{D}$, and $\tens{Q}^*$ are a sufficient set of dependencies for the learning task. From this perspective, the balance between the benefits of task simplification and implicit regularization seems to shift depending on the model inputs. With $\tens{Q}^*$, in the absence of overfitting, learning difficulty may matter more, whereas without $\tens{Q}^*$, models are prone to overfitting, so regularization may be more important.

\begin{figure}
    \centering
    \includegraphics[width=.90\textwidth]{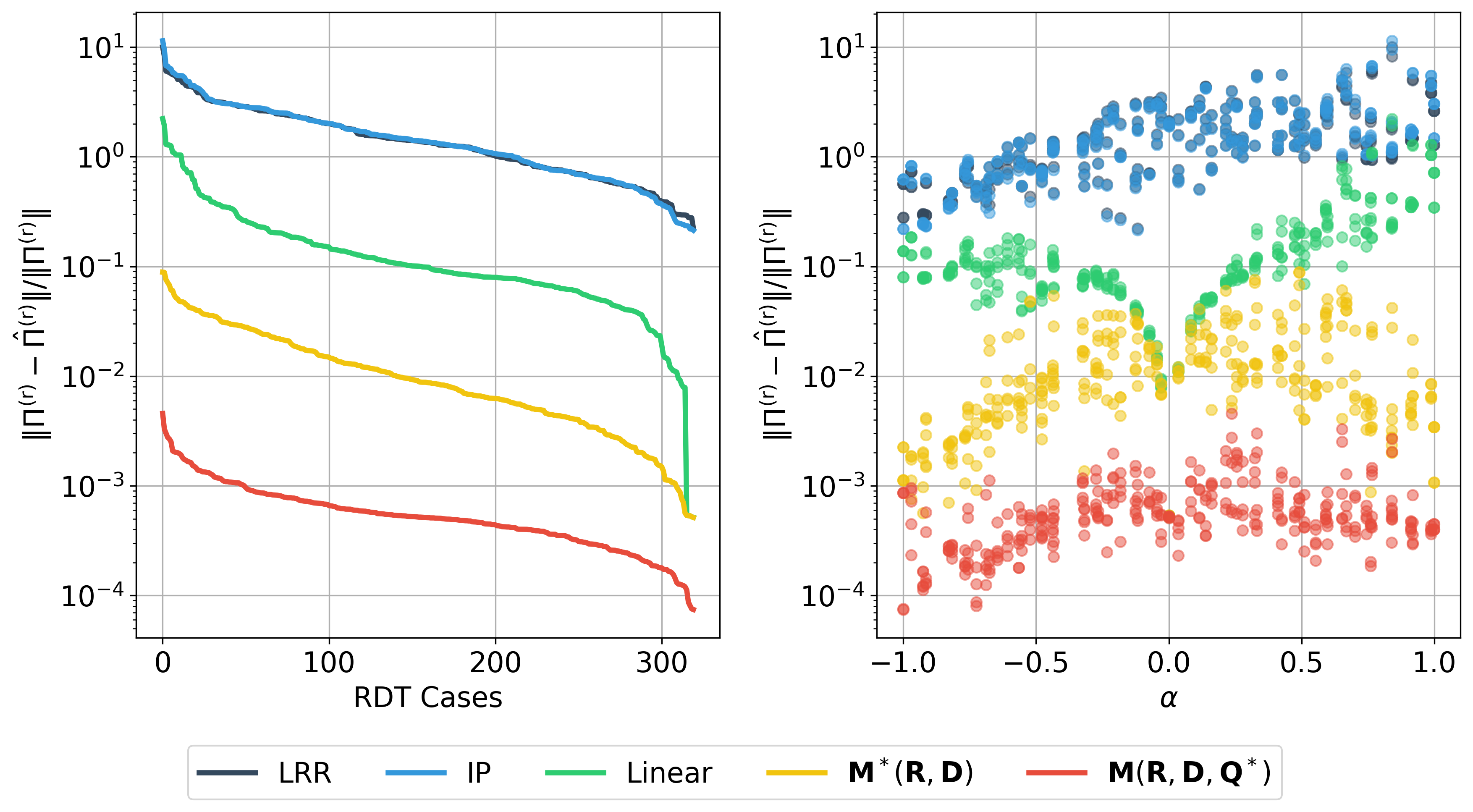}
    \caption{Error in predictions of the rapid pressure-rate-of-strain-tensor $\Pi_{ij}^{(r)}$. Left: Errors plotted in decreasing order. Right: Errors ordered by the magnitude of strain, with $\abs{\alpha} = 1$ representing pure strain and $\abs{\alpha} = 0$ representing pure rotation (A vertical slice thus represents the same velocity gradient for the right plot, but not the left). Shown are results for two commonly used models labeled LRR \citep{launder_progress_1975} and IP \citep{launder_second-moment_1989}, the linear structure tensor model proposed in \citep{kassinos_structure-based_1995}, and the two ENN models developed here.}
    \label{fig:m}
\end{figure}

Errors in the ENN models for $\tens{M}(\rst, \tens{D}, \tens{Q}^*)$ and $\tens{M}^*(\rst, \tens{D})$ were compared to those of the linear structure tensor model from \citep{kassinos_structure-based_1995} and two commonly used models: the Launder, Reece \& Rodi \citep{launder_progress_1975} (LRR)  model and the isotropisation of production \citep{launder_second-moment_1989} (IP) model (Fig.~\ref{fig:m}). To avoid presenting results that may suffer from overfitting, the $\tens{M}^*$ model trained on all components (without constraints) is used. For predicting $\tens{M}$, the model trained with constraints is used, as it performs better without overfitting. In Fig.~\ref{fig:m}, relative error in the rapid pressure-strain tensor $\Pi_{ij}^{(r)}$ in the Reynolds stress transport equation (Eq.~\ref{eq:RSeq}) is plotted, since it can be expressed in terms of $\tens{M}$ by Eq.~\ref{eq:rapidPi}. $\tens{M}$ also appears in the RDT evolution equation for $\tens{Q}$ (Eq.~\ref{eq:rdt-q}).

The three structure-tensor-based models (linear, $\tens{M}^*(\rst, \tens{D})$ and $\tens{M}(\rst, \tens{D}, \tens{Q}^*)$) represent a progression of nonlinear complexity. In particular, symmetries require that there can be no linear dependence of $\tens{M}^*$ on $\tens{Q}^*$ \citep{kassinos_structure-based_1995}, so when the ENN model for $\tens{M}^*(\rst, \tens{D})$ is used to determine $\tens{M}$ through Eq.~\ref{eq:m-decomp}, the resulting $\tens{M}$ model is
linear in $\tens{Q}^*$ and nonlinear in $\rst$ and $\tens{D}$. In contrast, the ENN $\tens{M}(\rst, \tens{D}, \tens{Q}^*)$ model is nonlinear in all three arguments. By comparing the performance of these three models, we can assess the importance of nonlinear dependence on $\rst$ and $\tens{D}$, and of nonlinear dependence on $\tens{Q}^*$.

The results in Fig.~\ref{fig:m} indicate that the linear model has order of magnitude smaller errors than the LRR and IP models, the $\tens{M}^*(\rst, \tens{D})$ model has errors that are order of magnitude smaller again; and finally, the $\tens{M}(\rst, \tens{D}, \tens{Q}^*)$ has yet another order of magnitude smaller error. The net effect being that the $\tens{M}(\rst, \tens{D}, \tens{Q}^*)$ model has three orders of magnitude smaller errors than the commonly used LRR and IP models, with relative errors in the pressure-strain term less than 0.5\% and most cases exhibiting errors less than 0.1\%. It appears then that nonlinearity in all three structure tensor arguments is important, and that the nonlinear ENN $\tens{M}(\rst, \tens{D}, \tens{Q}^*)$ model provides a highly accurate representation of the rapid pressure-strain term in the Reynolds stress transport equation, at least in the case of RDT. 

It turns out that when supervising only on the unconstrained spherical components, there is no distinction between learning $\tens{M}$ and $\tens{M}^*$. The reason is that, in both cases, the only components to be learned are those of the $\ell = 4$ irreducible representation, and these are identical for $\tens{M}$ and $\tens{M}^*$ provided that no term in Eq.~\ref{eq:m-decomp} other than $\tens{M}^*$ contains an $\ell = 4$ component.

Indeed, the $\ell = 4$ term of a fourth-order tensor decomposition corresponds to the fully symmetric trace-free (STF) component, to which none of the remaining terms in Eq.~\ref{eq:m-decomp} contribute. For the terms involving the Levi-Civita symbol, the contraction creates an antisymmetric pair of indices, which vanish under full symmetrization. For the others, the presence of at least one pair of free indices over the Kronecker delta means that these terms will generally have a nonzero trace and vanish in an STF projection. An alternative, representation theory perspective is that the terms other than $\tens{M}^*$ consist of either the Levi-Civita symbol or the Kronecker delta, which are $0\rm{o}$ and $0\rm{e}$, respectively, coupled with one other tensor having at most an $\ell = 3$ component, and hence the creation of an $\ell = 4$ component is precluded by selection rules in Sec.~\ref{sec:enns}.

\subsubsection[Modeling L]{Modeling $\tens{L}$}\label{sec:modeling-L}

The $\tens{L}$ tensor decomposes into irreducible representations $0\rm{e} + 2\rm{e} + 4\rm{e}$, the same as in $\tens{M}^*$. The supervision of all or only the unknown components of the irreducible representations was found to make a negligible difference when predicting $\tens{L}$, both as a function of $\tens{D}$ and as a function of $\tens{D}$ and $\tens{Q}^*$. The reported models were arbitrarily chosen to be those trained when supervising on only the unknowns.

\begin{figure}
    \centering
    \includegraphics[width=.9\textwidth]{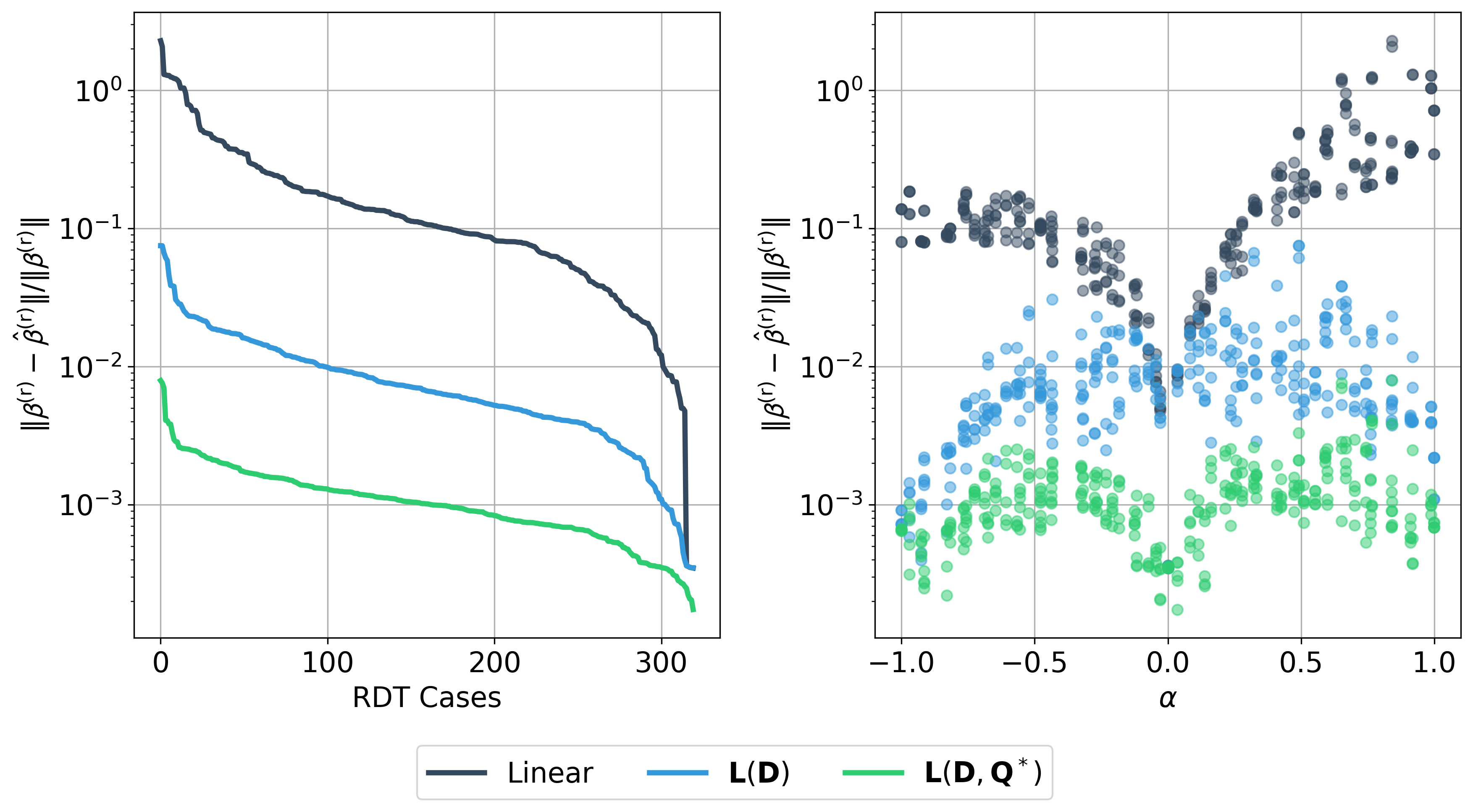}
    \caption{Error in predictions of $\beta_{ij}^{(r)}$. Left: Errors listed in decreasing order. Right. Errors ordered by amount of strain (See Fig.~\ref{fig:m}).}
    \label{fig:l}
\end{figure}

The error in predicting
\begin{equation}\label{eq:beta}
    \beta_{ij}^{(\mathrm{r})} \equiv 2\pdv{U_m}{x_n}\qty(L_{imnj} - M_{mnij}),
\end{equation}
which appears in the transport equation for $\tens{D}$, is shown in Fig.~\ref{fig:l} for the nonlinear $\tens{L}(\tens{D})$ and $\tens{L}(\tens{D}, \tens{Q}^*)$ models, as well as the linear model from~\citep{kassinos_one-point_2001}. For the $\tens{M}$ term in Eq.~\ref{eq:beta}, models from Sec.~\ref{sec:modeling-M} are used instead of the ground truth, reflecting the fact that closures for $\tens{L}$ and $\tens{M}$ must be used together in practice. The $\tens{L}(\tens{D})$ model is paired with the $\tens{M}^*(\tens{R}, \tens{D})$ model to limit $\tens{Q}^*$ dependence to the linear terms in Eq.~\ref{eq:m-decomp}. The $\tens{L}(\tens{D}, \tens{Q}^*)$ model is paired with the $\tens{M}(\tens{R}, \tens{D}, \tens{Q}^*)$ model and admits general nonlinear $\tens{Q}^*$ dependence.

The performance of $\tens{L}(\tens{D})$ and $\tens{L}(\tens{D}, \tens{Q}^*)$ models across a sweep of model configurations exhibits much less variation than that of $\tens{M}^*$, $\tens{M}$, and $\tens{J}$ models in sweeps over the same configurations. No overfitting is observed with or without $\tens{Q}^*$ dependence, regardless of whether constraints are enforced in training, and enforcing the constraints does not benefit the performance of models that depend on $\tens{Q}^*$. Hence, the differences in training dynamics affecting the $\tens{M}^*$ and $\tens{M}$ models do not seem to be present, or at least do not have the same level of influence, when modeling $\tens{L}$ (although they do for modeling $\tens{J}$, as will be mentioned in Sec.~\ref{sec:modeling-J}).

It can be seen that the nonlinear ENN structure tensor models outperform the linear model of \citep{kassinos_one-point_2001} across the range of mean velocity gradients. As with modeling $\tens{M}$, inclusion of general $\tens{Q}^*$ dependence leads to a clear performance improvement, with relative errors less than 1\% and often less than 0.1\%.

\subsubsection[Modeling J]{Modeling $\tens{J}$}\label{sec:modeling-J}

\begin{figure}
    \centering
    \includegraphics[width=.9\textwidth]{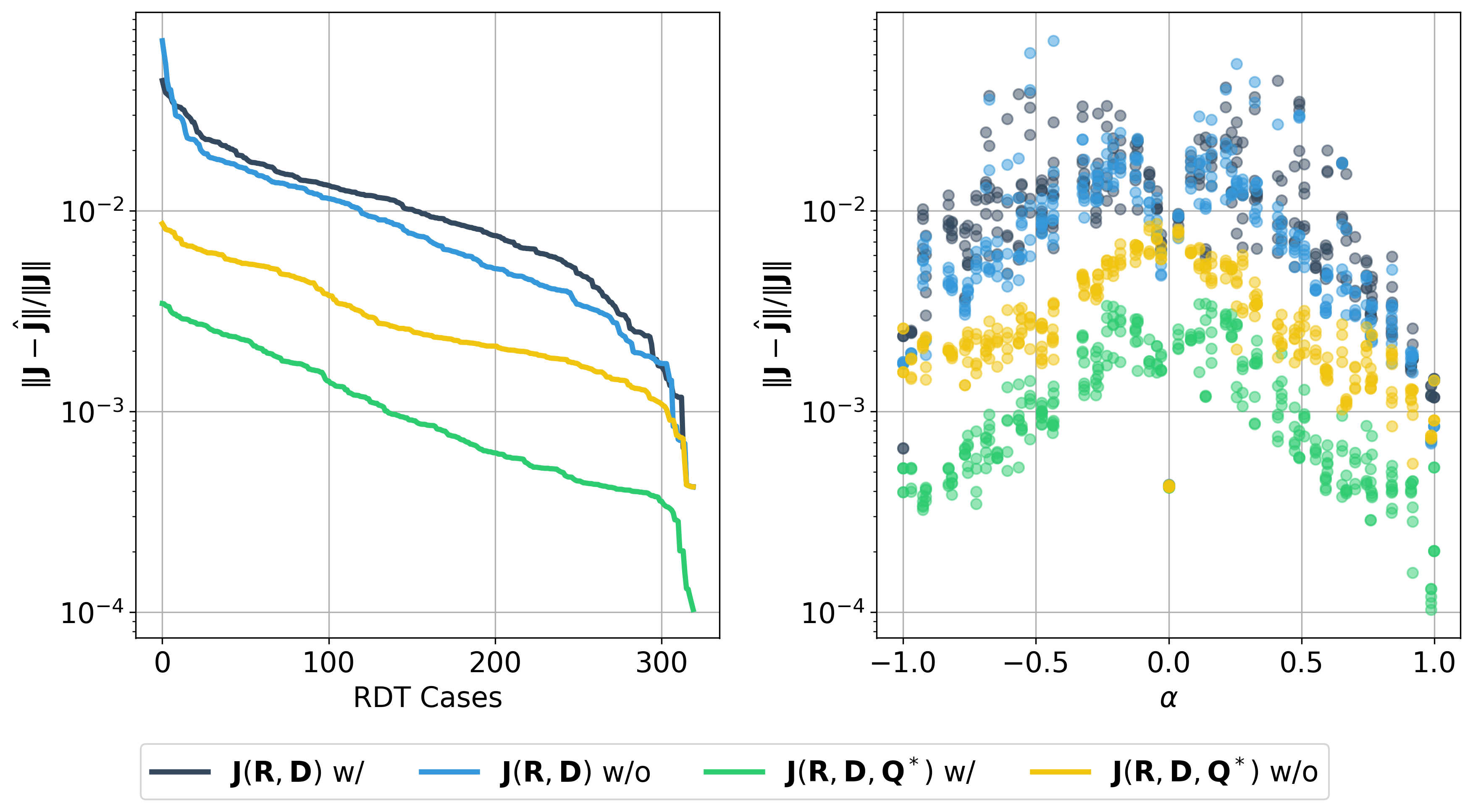}
    \caption{Error in predictions of $\tens{J}$. Left: Errors listed in decreasing order. Right. Errors ordered by amount of strain (See Fig.~\ref{fig:m}). In the legend, ``w/" and ``w/o" refer to training performed with and without the constraints in figure~\ref{fig:j-constraints} enforced, respectively.} 
    \label{fig:j}
\end{figure}

\begin{figure}
    \makebox[\textwidth][c]{%
        \begin{subfigure}[c]{0.45\textwidth}
            \centering
            \footnotesize
            \begin{tabular}{r@{${}={}$}l c}
                \toprule
                \multicolumn{2}{c}{Constraint} &
                \multicolumn{1}{c}{Irreps.} \\
                \midrule
                $J_{kkijr}$ & $0$ & \c{1o + 3o} \\
                $J_{ikjrk}$ & $0$ & \c{1o + 2o + 3o} \\
                $J_{kijrk}$ & $0$ & \c{1o + 2o + 3o} \\
                $\epsilon_{jsr}J_{sirkk}$ & $\rs_{ij}$ & \c{0o + 2o} \\
                $\epsilon_{irs}J_{srjkk}$ & $D_{ij}$ & \c{0o + 2o} \\
                $\sym(J_{ijrkk})$ & $Q^*_{ijr}$ & \c{1o + 3o} \\
                \bottomrule
            \end{tabular}
            \subcaption{Decomposition of linear constraints into irreducible representations.}
            \label{fig:j-constraints}
        \end{subfigure}
        \hspace{1em}%
        \begin{subfigure}[c]{0.5\textwidth}
            \centering
            \includegraphics[width=\linewidth]{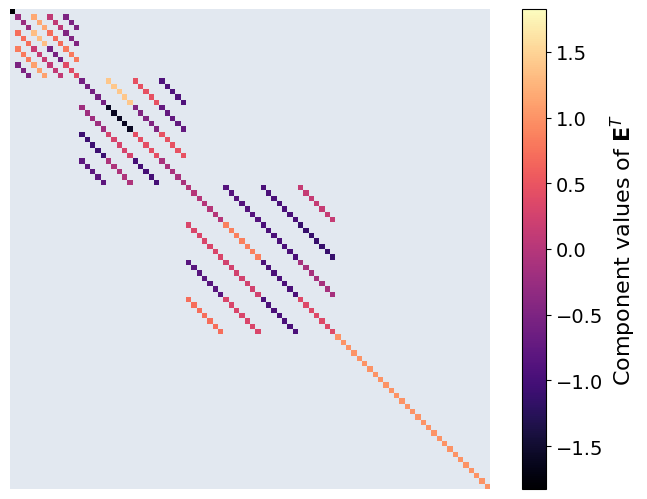}
            \subcaption{$\tens{E}^{\mathrm{T}}$ from Alg.~\ref{alg:lin-ch}, computed for decomposing the $\tens{J}$ tensor. All entries in gray are zero.}
            \label{fig:j-E}
        \end{subfigure}
    }
    \caption{The $\tens{J}$ tensor decomposes into $0\rm{o} + 4\times 1\rm{o} + 4\times 2\rm{o} + 4\times 3\rm{o} + 2\times 4\rm{o} + 5\rm{o}$. Of these, all the component values for irreducible representations with $\ell < 4$ are fixed by the constraints in Fig.~\ref{fig:j-constraints}. When the irreducible representations are ordered by increasing $\ell$, as in Fig.~\ref{fig:j-E}, $\tens{E}^{\mathrm{T}}$ is block diagonal, with each block representing an isotypic component. The subspaces for $\ell\in \{0, 1, 2, 3\}$ are those affected by the constraints.}
    \label{fig:E-matrix}
 \end{figure}

The $\tens{J}$ tensor decomposes into $0\rm{o} + 4\times 1\rm{o} + 4\times 2\rm{o} + 4\times 3\rm{o} + 2\times 4\rm{o} + 5\rm{o}$. The components for the irreducible representations with $\ell < 4$ are fixed by the constraints in figure~\ref{fig:j-constraints}, leaving the $2\times 4\rm{o} + 5\rm{o}$ components to be modeled. As $\tens{J}$ features the highest tensor order and largest number of constraints out of the three closure problems considered, we briefly use this case to illustrate the essence of Alg.~\ref{alg:lin-ch} in more detail.

When $\tens{S}_\pi\tens{P}_s$ is applied to $\vflat\qty(\tens{J})$, the Cartesian components of $\tens{J}$ are projected into bases that span irreducible subspaces. The particular bases are determined by the method used to solve Eq.~\ref{eq:infer-cob}. Meanwhile, the constraints on $\tens{J}$ can themselves be decomposed into irreducible subspaces, as shown in Fig.~\ref{fig:j-constraints}. While this decomposition is also achieved by solving Eq.~\ref{eq:infer-cob} and projecting with $\tens{S}_\pi\tens{P}_s$, the fact that the components of $\tens{J}$ are mixed by a contraction prior to this projection means that the resulting bases will not necessarily be the same. This is a reflection of the fact that isotypic component decompositions are not unique.

The new ENN layer discussed in Sec.~\ref{sec:lin-cc} corrects for this by using the $\tens{E}$ matrix, shown in Fig.~\ref{fig:j-E}, to synchronize the isotypic component decompositions. Since the loop in Alg.~\ref{alg:lin-ch} runs over $\ell$, the $\tens{E}$ matrix is sorted in ascending order of irreducible representations. The structure of the matrix is block diagonal as a consequence of Schur's lemma. Each block corresponds to a particular order $\ell$, and each block contains a number of constant matrices equal to the multiplicity that appears in the decomposition $0\rm{o} + 4\times 1\rm{o} + 4\times 2\rm{o} + 4\times 3\rm{o} + 2\times 4\rm{o} + 5\rm{o}$. That subspaces of a given $\ell$ are mixed with constant matrices is also a consequence of Schur's lemma, provided that the same basis convention is used for each individual subspace, as is the case here. The hierarchical block structure thus reflects the mixing freedom within isotypic components described in Sec.~\ref{sec:lin-cc}.

Modeling errors of ENN models for $\tens{J}$ for two different training strategies (with and without constraints) and model dependencies (with and without $\tens{Q}^*$) are compared in Fig.~\ref{fig:j}. Consistent with the discussion in Sec.~\ref{sec:modeling-M}, training with constraints leads to a clear performance improvement when $\tens{Q}^*$ is included as a model dependency, whereas omitting constraints leads to slightly better performance without $\tens{Q}^*$. Likewise, a sweep over model hyperparameters shows instances of overfitting only when $\tens{Q}^*$ is not a dependency. As with modeling $\tens{M}$ and $\tens{L}$, including $\tens{Q}^*$ dependence leads to errors roughly an order of magnitude lower than models based on $\tens{R}$ and $\tens{D}$ alone. The errors are of a similar magnitude to those in Figs.~\ref{fig:m} and \ref{fig:l} (less than 1\% relative error, and often less than 0.1\%), providing evidence that $\tens{R}$, $\tens{D}$, and $\tens{Q}^*$ are sufficient dependencies for this closure problem in addition to those for $\tens{M}$ and $\tens{L}$.

\section{Discussion}\label{sec:discussion}

In this paper, equivariant neural networks (ENNs) are used to represent high-order tensor functions of high-order tensor arguments, in this case applied to Reynolds-averaged Navier-Stokes (RANS) turbulence modeling using the structure tensors of Kassinos and Reynolds \citep{kassinos_one-point_2001}. When modeling the effects of the rapid pressure on the evolution of the structure tensors, fourth and fifth-order tensors must be expressed as functions of the structure tensors, but due to their definition, they also must satisfy a variety of index permutation symmetries and linear constraints. Using tools from tensor representation theory, a novel method was developed to incorporate the linear constraints into the architecture of the ENNs, along with equivariance and index symmetries. As a consequence, the resulting ENN representation satisfies equivariance, index symmetry, and applicable linear constraints exactly. This technique will also be valuable for tensor modeling in other applications, where symmetries and constraints beyond equivariance are common. In particular, it is applicable for modeling all RANS modeling closure terms. 

Kassinos and Reynolds were not able to test their hypothesis that the richer description of the anisotropy of turbulence provided by the structure tensors was sufficient to accurately close RANS models. At least in part, this was  because the analytical tools to develop the required tensor bases were not available. While recent advances \citep{olive_minimal_2017,olive2014isotropic} provide a path to perform such analysis, it is nonetheless arduous and time-consuming. So much so that we have not yet completed the analysis to formulate the structure tensor models for the rapid pressure terms in this way. The ENN formulation described here side-steps this analysis by building equivariance and other constraints into a neural network. It thus enables the testing of the Kassinos and Reynolds hypothesis, as we have done here for the rapid pressure terms. Because the ENN formulation is simpler to develop than the tensor basis formulation, it is particularly valuable for evaluating model dependencies, but at the expense of interpretability. It is thus still useful to pursue the development of tensor basis representations, which will allow assessment of what terms are important or negligible, and will enable principled approximations. Such tensor basis formulations are currently being developed.

To evaluate both the utility of the ENN formulation and the Kassinos \& Reynolds hypothesis, ENN models for the tensors appearing in the rapid pressure terms were formulated, trained, and tested (Sec.~\ref{sec:results}). The required data was obtained for a wide variety of specified mean velocity gradients using rapid distortion theory (RDT), as this is the simplest context in which the rapid pressure terms appear. The resulting models are remarkably successful at representing the rapid pressure terms. Relative errors for the ENN model with $\tens{R}$, $\tens{D}$ and $\tens{Q}^*$ as inputs are more than three orders of magnitude lower than the classical Launder, Reece \& Rodi (LRR) and isotropisation of production (IP) models \cite {launder_progress_1975,launder_second-moment_1989} for the term they model (the $\tens{M}$ tensor, see Sec. \ref{sec:modeling-M}). Relative errors in the rapid term for the ENN model of $\tens{M}$  are all less than 1\%, and in most cases less than 0.1\%. The same is true for the ENN models of the other two tensors appearing in the rapid pressure terms ($\tens{L}$ and $\tens{J}$, see Sec.~\ref{sec:modeling-L} and \ref{sec:modeling-J}).

Clearly, the ENN formulation developed here was successful in representing the tensor relationships underlying the dependence of the rapid pressure terms on the structure tensors. Specifying the ENN architecture requires truncation of the order of intermediate features retained in the Clebsch-Gordan tensor product, in this case at order $\ell_{\rm max}=6$, and the selection of other architecture characteristics. The choices made here result in modest sized ENNs (order 100k parameters, see table~\ref{tab:model-params}), which are tractable to train and evaluate. It appears that a higher-order truncation and more complex networks are not required.

The results in Sec.~\ref{sec:results} also validate the Kassinos \& Reynolds hypothesis as applied to the rapid pressure terms, at least for homogeneous turbulence in the RDT limit. That is, for these terms, in this context, the structure functions provide a sufficiently rich description of the turbulence anisotropy to allow accurate prediction of the rapid pressure terms. Furthermore, the results indicate that a nonlinear dependence on each of the three structure tensors ($\tens{R}$, $\tens{D}$ and $\tens{Q}^*$) is required. Linear models for $\tens{M}$ and $\tens{L}$ were formulated in \citep{kassinos_one-point_2001} and have errors that are two orders of magnitude larger than the nonlinear model for most cases. The importance of a nonlinear dependence on $\tens{Q}^*$ is also indicated by the order of magnitude larger errors incurred by models that are linear in $\tens{Q}^*$ and nonlinear in $\tens{R}$ and $\tens{D}$.  

The results reported here suggest several topics for further investigation. First, the ENN modeling paradigm can be applied to other terms that appear in the evolution equation for $\tens{Q}$. In homogeneous turbulence, these include the dissipation term, and the terms arising from the  Navier-Stokes nonlinear terms, including the so-called slow pressure terms. Second, the ENN models can be integrated with a solver for the $\tens{Q}$ evolution equation to solve the then closed RANS equations to assess the stability, realizability, and accuracy of the resulting solution. Finally, the computational efficiency of the ENNs should be improved to facilitate their use in flow solvers where the ENNs will have to be evaluated many times. In particular, there are opportunities for performance optimization of key equivariant operations like the Clebsch-Gordan tensor product.

\section*{Funding Statement}
The research reported here is supported by the U.S. Department of Energy, Office of Science, Office of Advanced Scientific Computing Research, Department of Energy Computational Science Graduate Fellowship under Award Number DE-SC0025528, and the National Science Foundation under Award Nos. CBET-2347422 \& CBET-2347423.

\section*{Competing Interests}
The author(s) declare none.

\section*{Acknowledgements}
Generative AI tools were used for sentence-level grammatical corrections and limited prose rewording during manuscript preparation.

\begin{appen}

\section{ENN architecture and training}\label{sec:train-details}

The backbone of all the equivariant MLPs is the \c{FullyConnectedTensorProduct} from the \c{e3nn} library. This module is initialized with \[\c{irrep\_normalization="norm"}\qq{and}\c{path\_normalization="path"},\] which insert a per-path multiplicative weight in Eq.~\ref{eq:cg-tp} based on the dimensions of the irreducible representations involved and the number of paths present (See the \c{e3nn} documentation at \url{https://docs.e3nn.org/}). Following \citep{brandstetter_geometric_2021}, biases are applied strictly to scalar outputs of the tensor product, and weights and biases are initialized based on the path fan-in for each output irreducible representation. An inverse rescaling is applied in the forward pass to compensate for the initialization. In the gating mechanism, the sigmoid linear unit (SiLU) and hyperbolic tangent (Tanh) activation functions are used for scalars of even and odd parity, respectively. All models were trained using the PyTorch library with the parameters in table~\ref{tab:hparams}.

\begin{table}
    \centering
    \begin{tabular}{rc}
        \toprule
        Parameter & Value \\
        \midrule
        Optimizer & Adam \\
        Epochs & 250 \\
        Batch size & 32 \\
        LR schedule & \c{OneCycleLR} \\
        Max LR & $5\times 10^{-3}$ \\
        Precision & float32 \\
        \bottomrule
    \end{tabular}
    \caption{Parameters used for ENN training.}
    \label{tab:hparams}
\end{table}

\begin{table}
    \centering
    \begin{tabular}{r|cc|cc|cc|cc|cc|cc}
        \toprule
        & \multicolumn{2}{c|}{$\tens{L}\qty(\tens{D})$} & \multicolumn{2}{c|}{$\tens{L}\qty(\tens{D}, \tens{Q}^*)$} & \multicolumn{2}{c|}{$\tens{M}^*\qty(\rst, \tens{D})$} & \multicolumn{2}{c|}{$\tens{M}\qty(\rst, \tens{D}, \tens{Q}^*)$} & \multicolumn{2}{c|}{$\tens{J}\qty(\rst, \tens{D})$} & \multicolumn{2}{c}{$\tens{J}\qty(\rst, \tens{D}, \tens{Q}^*)$}\\
        \midrule
        Constrained & \cmark & \xmark & \cmark & \xmark & \cmark & \xmark & \cmark & \xmark & \cmark & \xmark & \cmark & \xmark \\
        $\ell_{\max}$ & 6 & 6 & 6 & 6 & 6 & 6 & 6 & 6 & 6 & 6 & 6 & 6 \\
        Layers & 5 & 5 & 5 & 6 & 4 & 4 & 6 & 6 & 3 & 5 & 6 & 3 \\
        Hidden dim. & 128 & 256 & 128 & 256 & 256 & 256 & 512 & 256 & 512 & 256 & 256 & 256 \\
        Parameters & 19k & 28k & 22k & 45k & 42k & 42k & 198k & 81k & 65k & 56k & 80k & 36k \\
        \bottomrule
    \end{tabular}
    \caption{Parameters for models presented in Sec.~\ref{sec:results}.}
    \label{tab:model-params}
\end{table}

The parameters for the models presented in the main text are shown in table~\ref{tab:model-params}. Constrained refers to whether the model was supervised on all the irreducible representation components of the output tensor (unconstrained) or only on those not fixed by constraints (constrained). Hidden dimension refers to the total number of irreducible components in a given internal ENN layer. All internal layers share the same hidden dimension. The components in a hidden layer are distributed amongst all the irreducible representations up to and including $\ell_{\max}$, including both even and odd parity, such that the total number of components for each $\ell$ is approximately equal.

For a given column of table~\ref{tab:model-params}, the parameters correspond to the top-performing model identified in a sweep of 12 hyperparameter configurations, distinguished by the number of layers (3, 4, 5, or 6) and hidden dimension (128, 256, or 512). Sec.~\ref{sec:modeling-M} uses the unconstrained model for $\tens{M}^*$ and the constrained model for $\tens{M}$. Sec.~\ref{sec:modeling-L} uses the constrained models for both $\tens{L}(\tens{D})$ and $\tens{L}(\tens{D}, \tens{Q}^*)$. All $\tens{J}$ models in table~\ref{tab:model-params} appear in Sec.~\ref{sec:modeling-J}.

\section[Constraints on M*]{Constraints on $\tens{M}^*$}\label{sec:ms-constraints}

From table~\ref{tab:constraints}, $\tens{M}$ satisfies
\begin{equation}\label{eq:m-constraint}
    M_{ikkj} = 0,\quad M_{ijkk} = \rs_{ij},\quad M_{kkij} = D_{ij},\qq{and} \sym(\epsilon_{ipq}M_{jqpk}) = Q^*_{ijk}.
\end{equation}
The first three of these constraints are obtained by contracting two indices of $\tens{M}$. For $\tens{M}^*$, which is fully symmetric, any contraction of two indices produces the same second-order tensor, and the first three constraints for $\tens{M}$ collapse into one:
\begin{equation}\label{eq:ms-constraint}
    M^*_{ijkk} = \frac{1}{12}\sum_{\sigma\in S_4/S_2}M_{\sigma(i, j, k, k)} = \frac{1}{6}(\rs_{ij} + D_{ij}),
\end{equation}
where $S_4$ and $S_2$ are the symmetric groups on four and two elements, respectively. All of the terms in the sum are zero except for $M_{ijkk} = M_{jikk} = \rs_{ij}$ and $M_{kkij} = M_{kkji} = D_{ij}$ due to the symmetry of $\tens{M}$. Since $M^*_{ijkk}$ is a symmetric second-order tensor, Eq.~\ref{eq:ms-constraint} decomposes into $0\rm{e} + 2\rm{e}$ and represents six constraints.

As $\tens{M}^*$ decomposes into $0\rm{e} + 2\rm{e} + 4\rm{e}$, for there to remain an undetermined part of the tensor, it must be that the final constraint in Eq.~\ref{eq:m-constraint} does not impose any additional restrictions on $\tens{M}^*$. Indeed, writing $\tens{M} = \tens{M}^* + \tens{M}'$, where $\tens{M}'$ is the non-fully symmetric part of $\tens{M}$, the final constraint becomes
\begin{equation*}
    \sym[\epsilon_{ipq}(M'_{jqpk} + M^*_{jqpk})] = \sym(\epsilon_{ipq}M'_{jqpk}) + \sym(\epsilon_{ipq}M^*_{jqpk}) = Q^*_{ijk},
\end{equation*}
but for $\sym(\epsilon_{ipq}M^*_{jqpk})$, contracting a pair of symmetric and antisymmetric indices makes the term zero. Hence, the $\tens{Q}^*$ relation only constrains the non-fully symmetric part of $\tens{M}$, leaving the $\ell = 4$ components of $\tens{M}^*$ to be modeled.

\section[Invertibility of E]{Invertibility of $\tens{E}$}\label{sec:inv-E}

As explained in Sec.~\ref{sec:lin-cc}, the map from the spherical basis to the Cartesian basis is $\tens{P}_s^{\mathrm{T}}\tens{S}_{\pi}^{-1}(\tens{E}^{\mathrm{T}})^{-1}$, which involves the inverse of the $\tens{E}$ matrix from Alg.~\ref{alg:lin-ch}. The columns of the $\tens{E}$ matrix are populated by solutions to the least squares problem in Line 11 of the algorithm. In general, even if all the right-hand side vectors of a set of least squares problems are linearly independent, the least squares solutions may not be. However, in the algorithm, all of the solutions that populate $\tens{E}$ exactly satisfy the least squares problem.

Let the set of these solutions vectors be $\{\b{x}_1, \b{x}_2, \dots, \b{x}_k\}$. Writing 
\begin{equation*}
    c_1\b{x}_1 + c_2\b{x}_2 + \cdots + c_k\b{x}_k = \b{0}
\end{equation*}
and multiplying by $\tens{A}^{\mathrm{T}}$ from Alg.~\ref{alg:lin-ch},
\begin{equation*}
    \tens{A}^{\mathrm{T}}(c_1\b{x}_1 + c_2\b{x}_2 + \cdots + c_k\b{x}_k) = c_1\b{b}_1 + c_2\b{b}_2 + \cdots + c_k\b{b}_k = \b{0}.
\end{equation*}
The columns of $\tens{B}$ are linearly independent, so the only way for this equation to be true is if $c_1 = c_2 = \cdots = c_k = 0$. Hence, the vectors that populate the square matrix $\tens{E}$ are linearly independent, and the inverse exists.

\end{appen}\clearpage

\bibliographystyle{jfm}
\bibliography{jfm}

@BOOK{pope_turbulent_2000,
  AUTHOR = {Pope, S. B.},
  ADDRESS = {Cambridge ; New York},
  PUBLISHER = {Cambridge University Press},
  YEAR = {2000},
  ISBN = {978-0-521-59125-6 978-0-521-59886-6},
  TITLE = {Turbulent flows},
}

@ARTICLE{pope_more_1975,
  AUTHOR = {Pope, S. B.},
  LANGUAGE = {en},
  URL = {https://www.cambridge.org/core/journals/journal-of-fluid-mechanics/article/more-general-effectiveviscosity-hypothesis/86456F12CB23C8D2D9A2021CBB7FB732},
  YEAR = {1975},
  DOI = {10.1017/S0022112075003382},
  ISSN = {1469-7645, 0022-1120},
  JOURNAL = {Journal of Fluid Mechanics},
  NUMBER = {2},
  PAGES = {331--340},
  TITLE = {A more general effective-viscosity hypothesis},
  URLDATE = {2024-09-29},
  VOLUME = {72},
}

@INCOLLECTION{lumley_computational_1979,
  AUTHOR = {Lumley, John L.},
  EDITOR = {Yih, Chia-Shun},
  PUBLISHER = {Elsevier},
  URL = {https://www.sciencedirect.com/science/article/pii/S0065215608702667},
  BOOKTITLE = {Advances in {Applied} {Mechanics}},
  YEAR = {1979},
  DOI = {10.1016/S0065-2156(08)70266-7},
  PAGES = {123--176},
  TITLE = {Computational {Modeling} of {Turbulent} {Flows}*},
  URLDATE = {2025-03-10},
  VOLUME = {18},
}

@ARTICLE{speziale1987second,
  AUTHOR = {Speziale, Charles G},
  YEAR = {1987},
  JOURNAL = {Quarterly of applied mathematics},
  NUMBER = {4},
  PAGES = {721--733},
  TITLE = {Second-order closure models for rotating turbulent flows},
  VOLUME = {45},
}

@ARTICLE{launder_second-moment_1989,
  AUTHOR = {Launder, B. E.},
  LANGUAGE = {en},
  URL = {https://onlinelibrary.wiley.com/doi/abs/10.1002/fld.1650090806},
  YEAR = {1989},
  DOI = {10.1002/fld.1650090806},
  ISSN = {1097-0363},
  JOURNAL = {International Journal for Numerical Methods in Fluids},
  NUMBER = {8},
  PAGES = {963--985},
  TITLE = {Second-moment closure and its use in modelling turbulent industrial flows},
  URLDATE = {2025-04-28},
  VOLUME = {9},
}

@ARTICLE{speziale1991modelling,
  AUTHOR = {Speziale, Charles G and Sarkar, Sutanu and Gatski, Thomas B},
  PUBLISHER = {Cambridge University Press},
  YEAR = {1991},
  JOURNAL = {Journal of fluid mechanics},
  PAGES = {245--272},
  TITLE = {Modelling the pressure--strain correlation of turbulence: an invariant dynamical systems approach},
  VOLUME = {227},
}

@ARTICLE{so1991second,
  AUTHOR = {So, RMC and Lai, YG and Zhang, HS and Hwang, BC},
  YEAR = {1991},
  JOURNAL = {AIAA journal},
  NUMBER = {11},
  PAGES = {1819--1835},
  TITLE = {Second-order near-wall turbulence closures-A review},
  VOLUME = {29},
}

@ARTICLE{durbin1993reynolds,
  AUTHOR = {Durbin, Paul A},
  PUBLISHER = {Cambridge University Press},
  YEAR = {1993},
  JOURNAL = {Journal of Fluid Mechanics},
  PAGES = {465--498},
  TITLE = {A Reynolds stress model for near-wall turbulence},
  VOLUME = {249},
}

@ARTICLE{launder_progress_1975,
  AUTHOR = {Launder, B. E. and Reece, G. J. and Rodi, W.},
  LANGUAGE = {en},
  URL = {https://www.cambridge.org/core/journals/journal-of-fluid-mechanics/article/progress-in-the-development-of-a-reynoldsstress-turbulence-closure/796DDAC14EF54A84A36100565D3420D5},
  YEAR = {1975},
  DOI = {10.1017/S0022112075001814},
  ISSN = {1469-7645, 0022-1120},
  JOURNAL = {Journal of Fluid Mechanics},
  NUMBER = {3},
  PAGES = {537--566},
  TITLE = {Progress in the development of a {Reynolds}-stress turbulence closure},
  URLDATE = {2025-04-28},
  VOLUME = {68},
}

@BOOK{durbin2011statistical,
  AUTHOR = {Durbin, Paul A and Reif, BA Pettersson},
  PUBLISHER = {John Wiley \& Sons},
  YEAR = {2011},
  TITLE = {Statistical theory and modeling for turbulent flows},
}

@ARTICLE{kassinos_one-point_2001,
  AUTHOR = {Kassinos, S. C. and Reynolds, W. C. and Rogers, M. M.},
  LANGUAGE = {en},
  URL = {https://www.cambridge.org/core/product/identifier/S0022112000002615/type/journal_article},
  YEAR = {2001},
  DOI = {10.1017/S0022112000002615},
  ISSN = {0022-1120, 1469-7645},
  JOURNAL = {Journal of Fluid Mechanics},
  PAGES = {213--248},
  TITLE = {One-point turbulence structure tensors},
  URLDATE = {2025-01-06},
  VOLUME = {428},
}

@ARTICLE{hunt1990rapid,
  AUTHOR = {Hunt, Julian CR and Carruthers, David J},
  PUBLISHER = {Cambridge University Press},
  YEAR = {1990},
  JOURNAL = {Journal of Fluid Mechanics},
  PAGES = {497--532},
  TITLE = {Rapid distortion theory and the ‘problems’ of turbulence},
  VOLUME = {212},
}

@BOOK{lumley_stochastic_2007,
  AUTHOR = {Lumley, John L.},
  ADDRESS = {Mineola, N.Y},
  PUBLISHER = {Dover Publications},
  YEAR = {2007},
  EDITION = {Dover ed},
  ISBN = {978-0-486-46270-7},
  SERIES = {Dover books on engineering},
  NUMBER = {},
  TITLE = {Stochastic tools in turbulence},
}

@MISC{spalart_old-fashioned_2023,
  AUTHOR = {Spalart, Philippe},
  PUBLISHER = {arXiv},
  URL = {http://arxiv.org/abs/2308.00837},
  YEAR = {2023},
  DOI = {10.48550/arXiv.2308.00837},
  NOTE = {arXiv:2308.00837},
  TITLE = {An {Old}-{Fashioned} {Framework} for {Machine} {Learning} in {Turbulence} {Modeling}},
  URLDATE = {2024-10-20},
}

@ARTICLE{duraisamy_turbulence_2019,
  AUTHOR = {Duraisamy, Karthik and Iaccarino, Gianluca and Xiao, Heng},
  LANGUAGE = {en},
  URL = {https://www.annualreviews.org/content/journals/10.1146/annurev-fluid-010518-040547},
  YEAR = {2019},
  DOI = {10.1146/annurev-fluid-010518-040547},
  ISSN = {0066-4189, 1545-4479},
  JOURNAL = {Annual Review of Fluid Mechanics},
  NOTE = {Publisher: Annual Reviews},
  NUMBER = {Volume 51, 2019},
  PAGES = {357--377},
  TITLE = {Turbulence {Modeling} in the {Age} of {Data}},
  URLDATE = {2024-06-14},
  VOLUME = {51},
}

@ARTICLE{robertson_invariant_1940,
  AUTHOR = {Robertson, H. P.},
  LANGUAGE = {en},
  URL = {https://www.cambridge.org/core/journals/mathematical-proceedings-of-the-cambridge-philosophical-society/article/invariant-theory-of-isotropic-turbulence/20D2B827EAE1BA8EFDBD0AEB7069A059},
  YEAR = {1940},
  DOI = {10.1017/S0305004100017199},
  ISSN = {1469-8064, 0305-0041},
  JOURNAL = {Mathematical Proceedings of the Cambridge Philosophical Society},
  NUMBER = {2},
  PAGES = {209--223},
  TITLE = {The invariant theory of isotropic turbulence},
  URLDATE = {2025-03-03},
  VOLUME = {36},
}

@TECHREPORT{sarkar_simple_1989,
  AUTHOR = {Sarkar, Sutanu and Speziale, Charles G.},
  URL = {https://ntrs.nasa.gov/citations/19890011041},
  YEAR = {1989},
  NOTE = {NTRS Author Affiliations: Institute for Computer Applications in Science and Engineering NTRS Document ID: 19890011041 NTRS Research Center: Legacy CDMS (CDMS)},
  NUMBER = {NAS 1.26:181797},
  TITLE = {A simple nonlinear model for the return to isotropy in turbulence},
  TYPE = {techreport},
  URLDATE = {2025-03-06},
}

@ARTICLE{chung_nonlinear_1995,
  AUTHOR = {Chung, Myung Kyoon and Kim, Soong Kee},
  URL = {https://doi.org/10.1063/1.868760},
  YEAR = {1995},
  DOI = {10.1063/1.868760},
  ISSN = {1070-6631},
  JOURNAL = {Physics of Fluids},
  NUMBER = {6},
  PAGES = {1425--1437},
  TITLE = {A nonlinear return‐to‐isotropy model with {Reynolds} number and anisotropy dependency},
  URLDATE = {2025-03-12},
  VOLUME = {7},
}

@ARTICLE{ling_reynolds_2016,
  AUTHOR = {Ling, Julia and Kurzawski, Andrew and Templeton, Jeremy},
  LANGUAGE = {en},
  URL = {https://www.cambridge.org/core/journals/journal-of-fluid-mechanics/article/reynolds-averaged-turbulence-modelling-using-deep-neural-networks-with-embedded-invariance/0B280EEE89C74A7BF651C422F8FBD1EB},
  YEAR = {2016},
  DOI = {10.1017/jfm.2016.615},
  ISSN = {0022-1120, 1469-7645},
  JOURNAL = {Journal of Fluid Mechanics},
  PAGES = {155--166},
  TITLE = {Reynolds averaged turbulence modelling using deep neural networks with embedded invariance},
  URLDATE = {2024-09-29},
  VOLUME = {807},
}

@ARTICLE{kaandorp_data-driven_2020,
  AUTHOR = {Kaandorp, Mikael L. A. and Dwight, Richard P.},
  URL = {https://www.sciencedirect.com/science/article/pii/S0045793020300700},
  YEAR = {2020},
  DOI = {10.1016/j.compfluid.2020.104497},
  ISSN = {0045-7930},
  JOURNAL = {Computers \& Fluids},
  PAGES = {104497},
  TITLE = {Data-driven modelling of the {Reynolds} stress tensor using random forests with invariance},
  URLDATE = {2025-04-28},
  VOLUME = {202},
}

@ARTICLE{cai_revisiting_2024,
  AUTHOR = {Cai, Jiayi and Angeli, Pierre-Emmanuel and Martinez, Jean-Marc and Damblin, Guillaume and Lucor, Didier},
  URL = {http://arxiv.org/abs/2403.11746},
  YEAR = {2024},
  DOI = {10.1016/j.compfluid.2024.106246},
  ISSN = {00457930},
  JOURNAL = {Computers \& Fluids},
  NOTE = {arXiv:2403.11746 [physics]},
  PAGES = {106246},
  SHORTTITLE = {Revisiting {Tensor} {Basis} {Neural} {Networks} for {Reynolds} stress modeling},
  TITLE = {Revisiting {Tensor} {Basis} {Neural} {Networks} for {Reynolds} stress modeling: application to plane channel and square duct flows},
  URLDATE = {2025-04-28},
  VOLUME = {275},
}

@ARTICLE{spencer_theory_1958,
  AUTHOR = {Spencer, A. J. M. and Rivlin, R. S.},
  LANGUAGE = {en},
  URL = {https://doi.org/10.1007/BF00277933},
  YEAR = {1958},
  DOI = {10.1007/BF00277933},
  ISSN = {1432-0673},
  JOURNAL = {Archive for Rational Mechanics and Analysis},
  NUMBER = {1},
  PAGES = {309--336},
  TITLE = {The theory of matrix polynomials and its application to the mechanics of isotropic continua},
  URLDATE = {2025-09-03},
  VOLUME = {2},
}

@ARTICLE{zhang_ensemble_2022,
  AUTHOR = {Zhang, Xin-Lei and Xiao, Heng and Luo, Xiaodong and He, Guowei},
  LANGUAGE = {en},
  URL = {https://www.cambridge.org/core/journals/journal-of-fluid-mechanics/article/ensemble-kalman-method-for-learning-turbulence-models-from-indirect-observation-data/1017FA12DF01850C1579ABCFB09405E2},
  YEAR = {2022},
  DOI = {10.1017/jfm.2022.744},
  ISSN = {0022-1120, 1469-7645},
  JOURNAL = {Journal of Fluid Mechanics},
  PAGES = {A26},
  TITLE = {Ensemble {Kalman} method for learning turbulence models from indirect observation data},
  URLDATE = {2025-09-03},
  VOLUME = {949},
}

@ARTICLE{lennon_scientific_2023,
  AUTHOR = {Lennon, Kyle R. and McKinley, Gareth H. and Swan, James W.},
  URL = {https://www.pnas.org/doi/10.1073/pnas.2304669120},
  YEAR = {2023},
  DOI = {10.1073/pnas.2304669120},
  JOURNAL = {Proceedings of the National Academy of Sciences},
  NOTE = {Publisher: Proceedings of the National Academy of Sciences},
  NUMBER = {27},
  PAGES = {e2304669120},
  TITLE = {Scientific machine learning for modeling and simulating complex fluids},
  URLDATE = {2025-09-03},
  VOLUME = {120},
}

@INCOLLECTION{parmar_generalized_2020,
  AUTHOR = {Parmar, Basu and Peters, Eric and Jansen, Kenneth E. and Doostan, Alireza and Evans, John A.},
  PUBLISHER = {American Institute of Aeronautics and Astronautics},
  URL = {https://arc.aiaa.org/doi/10.2514/6.2020-0351},
  BOOKTITLE = {{AIAA} {Scitech} 2020 {Forum}},
  YEAR = {2020},
  DOI = {10.2514/6.2020-0351},
  SERIES = {{AIAA} {SciTech} {Forum}},
  NUMBER = {},
  TITLE = {Generalized {Non}-{Linear} {Eddy} {Viscosity} {Models} for {Data}-{Assisted} {Reynolds} {Stress} {Closure}},
  URLDATE = {2025-09-03},
}

@MISC{sunol_learning_2025,
  AUTHOR = {Sunol, Alp M. and Roggeveen, James V. and Alhashim, Mohammed G. and Bae, Henry S. and Brenner, Michael P.},
  PUBLISHER = {arXiv},
  URL = {http://arxiv.org/abs/2510.24673},
  YEAR = {2025},
  DOI = {10.48550/arXiv.2510.24673},
  NOTE = {arXiv:2510.24673 [physics]},
  TITLE = {Learning constitutive models and rheology from partial flow measurements},
  URLDATE = {2025-11-09},
}

@ARTICLE{smith_isotropic_1971,
  AUTHOR = {Smith, G. F.},
  URL = {https://www.sciencedirect.com/science/article/pii/0020722571900231},
  YEAR = {1971},
  DOI = {10.1016/0020-7225(71)90023-1},
  ISSN = {0020-7225},
  JOURNAL = {International Journal of Engineering Science},
  NUMBER = {10},
  PAGES = {899--916},
  TITLE = {On isotropic functions of symmetric tensors, skew-symmetric tensors and vectors},
  URLDATE = {2025-05-05},
  VOLUME = {9},
}

@ARTICLE{pennisi_third_1992,
  AUTHOR = {Pennisi, S.},
  URL = {https://www.sciencedirect.com/science/article/pii/0020722592900115},
  YEAR = {1992},
  DOI = {10.1016/0020-7225(92)90011-5},
  ISSN = {0020-7225},
  JOURNAL = {International Journal of Engineering Science},
  NUMBER = {5},
  PAGES = {679--692},
  TITLE = {On third order tensor-valued isotropic functions},
  URLDATE = {2025-05-07},
  VOLUME = {30},
}

@ARTICLE{prakash_invariant_2022,
  AUTHOR = {Prakash, Aviral and Jansen, Kenneth E. and Evans, John A.},
  URL = {https://www.sciencedirect.com/science/article/pii/S0045782522004923},
  YEAR = {2022},
  DOI = {10.1016/j.cma.2022.115457},
  ISSN = {0045-7825},
  JOURNAL = {Computer Methods in Applied Mechanics and Engineering},
  PAGES = {115457},
  TITLE = {Invariant data-driven subgrid stress modeling in the strain-rate eigenframe for large eddy simulation},
  URLDATE = {2025-09-03},
  VOLUME = {399},
}

@MISC{duval_hitchhikers_2024,
  AUTHOR = {Duval, Alexandre and Mathis, Simon V. and Joshi, Chaitanya K. and Schmidt, Victor and Miret, Santiago and Malliaros, Fragkiskos D. and Cohen, Taco and Liò, Pietro and Bengio, Yoshua and Bronstein, Michael},
  PUBLISHER = {arXiv},
  URL = {http://arxiv.org/abs/2312.07511},
  YEAR = {2024},
  DOI = {10.48550/arXiv.2312.07511},
  NOTE = {arXiv:2312.07511 [cs]},
  TITLE = {A {Hitchhiker}'s {Guide} to {Geometric} {GNNs} for {3D} {Atomic} {Systems}},
  URLDATE = {2025-01-10},
}

@MISC{schutt_schnet_2017,
  AUTHOR = {Schütt, Kristof T. and Kindermans, Pieter-Jan and Sauceda, Huziel E. and Chmiela, Stefan and Tkatchenko, Alexandre and Müller, Klaus-Robert},
  PUBLISHER = {arXiv},
  URL = {http://arxiv.org/abs/1706.08566},
  YEAR = {2017},
  DOI = {10.48550/arXiv.1706.08566},
  NOTE = {arXiv:1706.08566 [stat]},
  SHORTTITLE = {{SchNet}},
  TITLE = {{SchNet}: {A} continuous-filter convolutional neural network for modeling quantum interactions},
  URLDATE = {2025-09-03},
}

@MISC{gasteiger_directional_2022,
  AUTHOR = {Gasteiger, Johannes and Groß, Janek and Günnemann, Stephan},
  PUBLISHER = {arXiv},
  URL = {http://arxiv.org/abs/2003.03123},
  YEAR = {2022},
  DOI = {10.48550/arXiv.2003.03123},
  NOTE = {arXiv:2003.03123 [cs]},
  TITLE = {Directional {Message} {Passing} for {Molecular} {Graphs}},
  URLDATE = {2025-09-03},
}

@MISC{thomas_tensor_2018,
  AUTHOR = {Thomas, Nathaniel and Smidt, Tess and Kearnes, Steven and Yang, Lusann and Li, Li and Kohlhoff, Kai and Riley, Patrick},
  LANGUAGE = {en},
  PUBLISHER = {arXiv},
  URL = {http://arxiv.org/abs/1802.08219},
  YEAR = {2018},
  NOTE = {arXiv:1802.08219 [cs]},
  SHORTTITLE = {Tensor field networks},
  TITLE = {Tensor field networks: {Rotation}- and translation-equivariant neural networks for {3D} point clouds},
  URLDATE = {2024-08-19},
}

@MISC{weiler_3d_2018,
  AUTHOR = {Weiler, Maurice and Geiger, Mario and Welling, Max and Boomsma, Wouter and Cohen, Taco},
  LANGUAGE = {en},
  PUBLISHER = {arXiv},
  URL = {http://arxiv.org/abs/1807.02547},
  YEAR = {2018},
  NOTE = {arXiv:1807.02547 [cs, stat]},
  SHORTTITLE = {{3D} {Steerable} {CNNs}},
  TITLE = {{3D} {Steerable} {CNNs}: {Learning} {Rotationally} {Equivariant} {Features} in {Volumetric} {Data}},
  URLDATE = {2024-08-14},
}

@INPROCEEDINGS{brandstetter_geometric_2021,
  AUTHOR = {Brandstetter, Johannes and Hesselink, Rob and Pol, Elise van der and Bekkers, Erik J. and Welling, Max},
  LANGUAGE = {en},
  URL = {https://openreview.net/forum?id=_xwr8gOBeV1},
  YEAR = {2021},
  BOOKTITLE = {International Conference on Learning Representations},
  TITLE = {Geometric and {Physical} {Quantities} improve {E}(3) {Equivariant} {Message} {Passing}},
  URLDATE = {2024-08-12},
}

@MISC{kondor_clebsch-gordan_2018,
  AUTHOR = {Kondor, Risi and Lin, Zhen and Trivedi, Shubhendu},
  PUBLISHER = {arXiv},
  URL = {http://arxiv.org/abs/1806.09231},
  YEAR = {2018},
  DOI = {10.48550/arXiv.1806.09231},
  NOTE = {arXiv:1806.09231 [stat] version: 2},
  SHORTTITLE = {Clebsch-{Gordan} {Nets}},
  TITLE = {Clebsch-{Gordan} {Nets}: a {Fully} {Fourier} {Space} {Spherical} {Convolutional} {Neural} {Network}},
  URLDATE = {2025-04-26},
}

@ARTICLE{jumper_highly_2021,
  AUTHOR = {Jumper, John and Evans, Richard and Pritzel, Alexander and Green, Tim and Figurnov, Michael and Ronneberger, Olaf and Tunyasuvunakool, Kathryn and Bates, Russ and Žídek, Augustin and Potapenko, Anna and Bridgland, Alex and Meyer, Clemens and Kohl, Simon A. A. and Ballard, Andrew J. and Cowie, Andrew and Romera-Paredes, Bernardino and Nikolov, Stanislav and Jain, Rishub and Adler, Jonas and Back, Trevor and Petersen, Stig and Reiman, David and Clancy, Ellen and Zielinski, Michal and Steinegger, Martin and Pacholska, Michalina and Berghammer, Tamas and Bodenstein, Sebastian and Silver, David and Vinyals, Oriol and Senior, Andrew W. and Kavukcuoglu, Koray and Kohli, Pushmeet and Hassabis, Demis},
  LANGUAGE = {en},
  URL = {https://www.nature.com/articles/s41586-021-03819-2},
  YEAR = {2021},
  DOI = {10.1038/s41586-021-03819-2},
  ISSN = {1476-4687},
  JOURNAL = {Nature},
  NOTE = {Publisher: Nature Publishing Group},
  NUMBER = {7873},
  PAGES = {583--589},
  TITLE = {Highly accurate protein structure prediction with {AlphaFold}},
  URLDATE = {2025-04-28},
  VOLUME = {596},
}

@ARTICLE{batzner_e3-equivariant_2022,
  AUTHOR = {Batzner, Simon and Musaelian, Albert and Sun, Lixin and Geiger, Mario and Mailoa, Jonathan P. and Kornbluth, Mordechai and Molinari, Nicola and Smidt, Tess E. and Kozinsky, Boris},
  LANGUAGE = {en},
  URL = {https://www.nature.com/articles/s41467-022-29939-5},
  YEAR = {2022},
  DOI = {10.1038/s41467-022-29939-5},
  ISSN = {2041-1723},
  JOURNAL = {Nature Communications},
  NOTE = {Publisher: Nature Publishing Group},
  NUMBER = {1},
  PAGES = {2453},
  TITLE = {E(3)-equivariant graph neural networks for data-efficient and accurate interatomic potentials},
  URLDATE = {2025-04-28},
  VOLUME = {13},
}

@MISC{zitnick_introduction_2020,
  AUTHOR = {Zitnick, C. Lawrence and Chanussot, Lowik and Das, Abhishek and Goyal, Siddharth and Heras-Domingo, Javier and Ho, Caleb and Hu, Weihua and Lavril, Thibaut and Palizhati, Aini and Riviere, Morgane and Shuaibi, Muhammed and Sriram, Anuroop and Tran, Kevin and Wood, Brandon and Yoon, Junwoong and Parikh, Devi and Ulissi, Zachary},
  PUBLISHER = {arXiv},
  URL = {http://arxiv.org/abs/2010.09435},
  YEAR = {2020},
  DOI = {10.48550/arXiv.2010.09435},
  NOTE = {arXiv:2010.09435 [cond-mat]},
  TITLE = {An {Introduction} to {Electrocatalyst} {Design} using {Machine} {Learning} for {Renewable} {Energy} {Storage}},
  URLDATE = {2025-04-28},
}

@ARTICLE{dauparas_robust_2022,
  AUTHOR = {Dauparas, J. and Anishchenko, I. and Bennett, N. and Bai, H. and Ragotte, R. J. and Milles, L. F. and Wicky, B. I. M. and Courbet, A. and de Haas, R. J. and Bethel, N. and Leung, P. J. Y. and Huddy, T. F. and Pellock, S. and Tischer, D. and Chan, F. and Koepnick, B. and Nguyen, H. and Kang, A. and Sankaran, B. and Bera, A. K. and King, N. P. and Baker, D.},
  URL = {https://www.science.org/doi/10.1126/science.add2187},
  YEAR = {2022},
  DOI = {10.1126/science.add2187},
  JOURNAL = {Science},
  NOTE = {Publisher: American Association for the Advancement of Science},
  NUMBER = {6615},
  PAGES = {49--56},
  TITLE = {Robust deep learning–based protein sequence design using {ProteinMPNN}},
  URLDATE = {2025-04-28},
  VOLUME = {378},
}

@ARTICLE{zhou_frame-independent_2022,
  AUTHOR = {Zhou, Xu-Hui and Han, Jiequn and Xiao, Heng},
  URL = {https://www.sciencedirect.com/science/article/pii/S0045782521005429},
  YEAR = {2022},
  DOI = {10.1016/j.cma.2021.114211},
  ISSN = {0045-7825},
  JOURNAL = {Computer Methods in Applied Mechanics and Engineering},
  PAGES = {114211},
  TITLE = {Frame-independent vector-cloud neural network for nonlocal constitutive modeling on arbitrary grids},
  URLDATE = {2025-09-03},
  VOLUME = {388},
}

@ARTICLE{han_equivariant_2023,
  AUTHOR = {Han, Jiequn and Zhou, Xu-Hui and Xiao, Heng},
  URL = {https://www.sciencedirect.com/science/article/pii/S0021999123003388},
  YEAR = {2023},
  DOI = {10.1016/j.jcp.2023.112243},
  ISSN = {0021-9991},
  JOURNAL = {Journal of Computational Physics},
  PAGES = {112243},
  TITLE = {An equivariant neural operator for developing nonlocal tensorial constitutive models},
  URLDATE = {2024-06-03},
  VOLUME = {488},
}

@ARTICLE{gao_roteqnet_2022,
  AUTHOR = {Gao, Liyao and Du, Yifan and Li, Hongshan and Lin, Guang},
  URL = {https://www.sciencedirect.com/science/article/pii/S0021999122002674},
  YEAR = {2022},
  DOI = {10.1016/j.jcp.2022.111205},
  ISSN = {0021-9991},
  JOURNAL = {Journal of Computational Physics},
  PAGES = {111205},
  SHORTTITLE = {{RotEqNet}},
  TITLE = {{RotEqNet}: {Rotation}-equivariant network for fluid systems with symmetric high-order tensors},
  URLDATE = {2025-09-03},
  VOLUME = {461},
}

@ARTICLE{kaszuba_implicit_2025,
  AUTHOR = {Kaszuba, Grzegorz and Krakowski, Tomasz and Ziegler, Bartosz and Jaszkiewicz, Andrzej and Sankowski, Piotr},
  URL = {https://doi.org/10.1063/5.0249490},
  YEAR = {2025},
  DOI = {10.1063/5.0249490},
  ISSN = {1070-6631},
  JOURNAL = {Physics of Fluids},
  NUMBER = {2},
  PAGES = {025137},
  TITLE = {Implicit modeling of equivariant tensor basis with {Euclidean} turbulence closure neural network},
  URLDATE = {2025-06-06},
  VOLUME = {37},
}

@MISC{pfaff_learning_2021,
  AUTHOR = {Pfaff, Tobias and Fortunato, Meire and Sanchez-Gonzalez, Alvaro and Battaglia, Peter W.},
  PUBLISHER = {arXiv},
  URL = {http://arxiv.org/abs/2010.03409},
  YEAR = {2021},
  NOTE = {arXiv:2010.03409 [cs]},
  TITLE = {Learning {Mesh}-{Based} {Simulation} with {Graph} {Networks}},
  URLDATE = {2024-03-04},
}

@MISC{toshev_learning_2023,
  AUTHOR = {Toshev, Artur P. and Galletti, Gianluca and Brandstetter, Johannes and Adami, Stefan and Adams, Nikolaus A.},
  PUBLISHER = {arXiv},
  URL = {http://arxiv.org/abs/2305.15603},
  YEAR = {2023},
  DOI = {10.48550/arXiv.2305.15603},
  NOTE = {arXiv:2305.15603 [cs]},
  TITLE = {Learning {Lagrangian} {Fluid} {Mechanics} with {E}(\$3\$)-{Equivariant} {Graph} {Neural} {Networks}},
  URLDATE = {2025-09-03},
}

@ARTICLE{list_rotational_2025,
  AUTHOR = {List, B. and Lino, M. and Thuerey, N.},
  URL = {https://doi.org/10.1063/5.0279499},
  YEAR = {2025},
  DOI = {10.1063/5.0279499},
  ISSN = {1070-6631},
  JOURNAL = {Physics of Fluids},
  NUMBER = {8},
  PAGES = {087178},
  TITLE = {Rotational equivariant graph neural networks via local eigenbasis transformations},
  URLDATE = {2025-08-22},
  VOLUME = {37},
}

@ARTICLE{lino_multi-scale_2022,
  AUTHOR = {Lino, Mario and Fotiadis, Stathi and Bharath, Anil A. and Cantwell, Chris D.},
  LANGUAGE = {en},
  URL = {https://pubs.aip.org/pof/article/34/8/087110/2847850/Multi-scale-rotation-equivariant-graph-neural},
  YEAR = {2022},
  DOI = {10.1063/5.0097679},
  ISSN = {1070-6631, 1089-7666},
  JOURNAL = {Physics of Fluids},
  NUMBER = {8},
  PAGES = {087110},
  TITLE = {Multi-scale rotation-equivariant graph neural networks for unsteady {Eulerian} fluid dynamics},
  URLDATE = {2025-09-02},
  VOLUME = {34},
}

@ARTICLE{milano_neural_2002,
  AUTHOR = {Milano, Michele and Koumoutsakos, Petros},
  URL = {https://www.sciencedirect.com/science/article/pii/S0021999102971469},
  YEAR = {2002},
  DOI = {10.1006/jcph.2002.7146},
  ISSN = {0021-9991},
  JOURNAL = {Journal of Computational Physics},
  NUMBER = {1},
  PAGES = {1--26},
  TITLE = {Neural {Network} {Modeling} for {Near} {Wall} {Turbulent} {Flow}},
  URLDATE = {2025-11-09},
  VOLUME = {182},
}

@ARTICLE{raissi_physics-informed_2019,
  AUTHOR = {Raissi, M. and Perdikaris, P. and Karniadakis, G.E.},
  LANGUAGE = {en},
  URL = {https://linkinghub.elsevier.com/retrieve/pii/S0021999118307125},
  YEAR = {2019},
  DOI = {10.1016/j.jcp.2018.10.045},
  ISSN = {00219991},
  JOURNAL = {Journal of Computational Physics},
  PAGES = {686--707},
  SHORTTITLE = {Physics-informed neural networks},
  TITLE = {Physics-informed neural networks: {A} deep learning framework for solving forward and inverse problems involving nonlinear partial differential equations},
  URLDATE = {2023-09-02},
  VOLUME = {378},
}

@ARTICLE{karniadakis_physics-informed_2021,
  AUTHOR = {Karniadakis, George Em and Kevrekidis, Ioannis G. and Lu, Lu and Perdikaris, Paris and Wang, Sifan and Yang, Liu},
  LANGUAGE = {en},
  URL = {https://www.nature.com/articles/s42254-021-00314-5},
  YEAR = {2021},
  DOI = {10.1038/s42254-021-00314-5},
  ISSN = {2522-5820},
  JOURNAL = {Nature Reviews Physics},
  NOTE = {Publisher: Nature Publishing Group},
  NUMBER = {6},
  PAGES = {422--440},
  TITLE = {Physics-informed machine learning},
  URLDATE = {2024-07-22},
  VOLUME = {3},
}

@MISC{wang_incorporating_2021,
  AUTHOR = {Wang, Rui and Walters, Robin and Yu, Rose},
  PUBLISHER = {arXiv},
  URL = {http://arxiv.org/abs/2002.03061},
  YEAR = {2021},
  DOI = {10.48550/arXiv.2002.03061},
  NOTE = {arXiv:2002.03061 [cs]},
  TITLE = {Incorporating {Symmetry} into {Deep} {Dynamics} {Models} for {Improved} {Generalization}},
  URLDATE = {2025-09-06},
}

@MISC{chen_group-theoretic_nodate,
  AUTHOR = {Chen, Shuxiao and Dobriban, Edgar and Lee, Jane H},
  LANGUAGE = {en},
  YEAR = {2019},
  NOTE = {arXiv preprint arXiv:1907.10905},
  TITLE = {A {Group}-{Theoretic} {Framework} for {Data} {Augmentation}},
}

@INPROCEEDINGS{liu_harnessing_2024,
  AUTHOR = {Liu, Ning and Fan, Yiming and Zeng, Xianyi and Klöwer, Milan and Zhang, Lu and Yu, Yue},
  LANGUAGE = {en},
  PUBLISHER = {PMLR},
  URL = {https://proceedings.mlr.press/v235/liu24p.html},
  BOOKTITLE = {Proceedings of the 41st {International} {Conference} on {Machine} {Learning}},
  YEAR = {2024},
  NOTE = {ISSN: 2640-3498},
  PAGES = {30965--30997},
  TITLE = {Harnessing the {Power} of {Neural} {Operators} with {Automatically} {Encoded} {Conservation} {Laws}},
  URLDATE = {2024-07-24},
}

@MISC{richter-powell_neural_2022,
  AUTHOR = {Richter-Powell, Jack and Lipman, Yaron and Chen, Ricky T. Q.},
  PUBLISHER = {arXiv},
  URL = {http://arxiv.org/abs/2210.01741},
  YEAR = {2022},
  DOI = {10.48550/arXiv.2210.01741},
  NOTE = {arXiv:2210.01741 [cs]},
  SHORTTITLE = {Neural {Conservation} {Laws}},
  TITLE = {Neural {Conservation} {Laws}: {A} {Divergence}-{Free} {Perspective}},
  URLDATE = {2025-09-06},
}

@MISC{chalapathi_scaling_2024,
  AUTHOR = {Chalapathi, Nithin and Du, Yiheng and Krishnapriyan, Aditi},
  PUBLISHER = {arXiv},
  URL = {http://arxiv.org/abs/2402.13412},
  YEAR = {2024},
  DOI = {10.48550/arXiv.2402.13412},
  NOTE = {arXiv:2402.13412 [cs]},
  TITLE = {Scaling physics-informed hard constraints with mixture-of-experts},
  URLDATE = {2025-09-06},
}

@MISC{vasilache_tensor_2018,
  AUTHOR = {Vasilache, Nicolas and Zinenko, Oleksandr and Theodoridis, Theodoros and Goyal, Priya and DeVito, Zachary and Moses, William S. and Verdoolaege, Sven and Adams, Andrew and Cohen, Albert},
  PUBLISHER = {arXiv},
  URL = {http://arxiv.org/abs/1802.04730},
  YEAR = {2018},
  DOI = {10.48550/arXiv.1802.04730},
  NOTE = {arXiv:1802.04730 [cs]},
  SHORTTITLE = {Tensor {Comprehensions}},
  TITLE = {Tensor {Comprehensions}: {Framework}-{Agnostic} {High}-{Performance} {Machine} {Learning} {Abstractions}},
  URLDATE = {2025-11-09},
}

@ARTICLE{johansson1994modelling,
  AUTHOR = {Johansson, Arne V and Hallbäck, Magnus},
  PUBLISHER = {Cambridge University Press},
  YEAR = {1994},
  JOURNAL = {Journal of Fluid Mechanics},
  PAGES = {143--168},
  TITLE = {Modelling of rapid pressure—strain in Reynolds-stress closures},
  VOLUME = {269},
}

@ARTICLE{girimaji2000pressure,
  AUTHOR = {Girimaji, Sharath S},
  PUBLISHER = {Cambridge University Press},
  YEAR = {2000},
  JOURNAL = {Journal of Fluid Mechanics},
  PAGES = {91--123},
  TITLE = {Pressure--strain correlation modelling of complex turbulent flows},
  VOLUME = {422},
}

@ARTICLE{mishra2010pressure,
  AUTHOR = {Mishra, Aashwin A and Girimaji, Sharath S},
  PUBLISHER = {Springer},
  YEAR = {2010},
  JOURNAL = {Flow, turbulence and combustion},
  NUMBER = {3},
  PAGES = {593--619},
  TITLE = {Pressure--strain correlation modeling: towards achieving consistency with rapid distortion theory},
  VOLUME = {85},
}

@ARTICLE{panda2020review,
  AUTHOR = {Panda, JP},
  PUBLISHER = {SAGE Publications Sage UK: London, England},
  YEAR = {2020},
  JOURNAL = {Proceedings of the Institution of Mechanical Engineers, Part C: Journal of Mechanical Engineering Science},
  NUMBER = {8},
  PAGES = {1528--1544},
  TITLE = {A review of pressure strain correlation modeling for Reynolds stress models},
  VOLUME = {234},
}

@PHDTHESIS{kassinos_structure-based_1995,
  AUTHOR = {Kassinos, Stavros Chistofer},
  SCHOOL = {Stanford University},
  LANGUAGE = {English},
  ADDRESS = {United States -- California},
  URL = {https://www.proquest.com/docview/304220375/abstract/E270CED471DA4FB6PQ/1},
  YEAR = {1995},
  NOTE = {ISBN: 9798209432388},
  TITLE = {A structure-based model for the rapid distortion of homogeneous turbulence},
  TYPE = {Ph.{D}.},
  URLDATE = {2025-03-19},
}

@BOOK{dresselhaus_group_2008,
  AUTHOR = {Dresselhaus, Mildred S. and Dresselhaus, Gene and Jorio, Ado},
  LANGUAGE = {eng},
  ADDRESS = {Berlin, Heidelberg},
  PUBLISHER = {Springer Berlin Heidelberg},
  YEAR = {2008},
  DOI = {10.1007/978-3-540-32899-5},
  ISBN = {978-3-540-32897-1 978-3-540-32899-5},
  SERIES = {{SpringerLink} {Bücher}},
  NUMBER = {},
  TITLE = {Group {Theory}},
}

@BOOK{georgi_lie_2018,
  AUTHOR = {Georgi, Howard},
  LANGUAGE = {eng},
  ADDRESS = {Boca Raton, Fl},
  PUBLISHER = {CRC Press},
  YEAR = {2018},
  DOI = {10.1201/9780429499210},
  EDITION = {Second edition},
  ISBN = {978-0-429-49921-0 978-0-429-96776-4},
  NUMBER = {54},
  SERIES = {Frontiers in physics},
  SHORTTITLE = {Lie algebras in particle physics},
  TITLE = {Lie algebras in particle physics: from isospin to unified theories},
}

@MISC{smidt_intuition_2023,
  AUTHOR = {Smidt, Tess},
  URL = {https://icml.cc/virtual/2023/28461},
  YEAR = {2023},
  NOTE = {Published: Keynote presentation at the 2nd Annual Workshop on Topology, Algebra, and Geometry in Machine Learning (TAG-ML) at ICML 2023},
  TITLE = {Intuition for the {Data} {Types} and {Interactions} of {Euclidean} {Neural} {Networks}},
}

@MISC{cohen_steerable_2016,
  AUTHOR = {Cohen, Taco S. and Welling, Max},
  PUBLISHER = {arXiv},
  URL = {http://arxiv.org/abs/1612.08498},
  YEAR = {2016},
  DOI = {10.48550/arXiv.1612.08498},
  NOTE = {arXiv:1612.08498 [cs]},
  TITLE = {Steerable {CNNs}},
  URLDATE = {2025-04-07},
}

@MISC{geiger_e3nn_2022,
  AUTHOR = {Geiger, Mario and Smidt, Tess},
  PUBLISHER = {arXiv},
  URL = {http://arxiv.org/abs/2207.09453},
  YEAR = {2022},
  DOI = {10.48550/arXiv.2207.09453},
  NOTE = {arXiv:2207.09453 [cs]},
  SHORTTITLE = {e3nn},
  TITLE = {e3nn: {Euclidean} {Neural} {Networks}},
  URLDATE = {2025-01-13},
}

@ARTICLE{sobol_distribution_1967,
  AUTHOR = {Sobol', I. M},
  URL = {https://www.sciencedirect.com/science/article/pii/0041555367901449},
  YEAR = {1967},
  DOI = {10.1016/0041-5553(67)90144-9},
  ISSN = {0041-5553},
  JOURNAL = {USSR Computational Mathematics and Mathematical Physics},
  NUMBER = {4},
  PAGES = {86--112},
  TITLE = {On the distribution of points in a cube and the approximate evaluation of integrals},
  URLDATE = {2025-09-10},
  VOLUME = {7},
}

@MISC{womersley_efficient_2017,
  AUTHOR = {Womersley, Robert S.},
  PUBLISHER = {arXiv},
  URL = {http://arxiv.org/abs/1709.01624},
  YEAR = {2017},
  DOI = {10.48550/arXiv.1709.01624},
  NOTE = {arXiv:1709.01624 [math]},
  TITLE = {Efficient {Spherical} {Designs} with {Good} {Geometric} {Properties}},
  URLDATE = {2025-07-14},
}

@ARTICLE{olive_minimal_2017,
  AUTHOR = {Olive, M. and Kolev, B. and Auffray, N.},
  LANGUAGE = {en},
  URL = {https://doi.org/10.1007/s00205-017-1127-y},
  YEAR = {2017},
  DOI = {10.1007/s00205-017-1127-y},
  ISSN = {1432-0673},
  JOURNAL = {Archive for Rational Mechanics and Analysis},
  NUMBER = {1},
  PAGES = {1--31},
  TITLE = {A {Minimal} {Integrity} {Basis} for the {Elasticity} {Tensor}},
  URLDATE = {2025-09-06},
  VOLUME = {226},
}

@ARTICLE{olive2014isotropic,
  AUTHOR = {Olive, Marc and Auffray, Nicolas},
  PUBLISHER = {AIP Publishing},
  YEAR = {2014},
  JOURNAL = {Journal of Mathematical Physics},
  NUMBER = {9},
  TITLE = {Isotropic invariants of a completely symmetric third-order tensor},
  VOLUME = {55},
}

\end{document}